\documentclass[onecolumn,10pt,showpacs,
amsmath,amssymb]{revtex4}
\usepackage{graphicx} 
\usepackage{amssymb, amsmath}
\usepackage{epsf}

\def\la{\; \raise0.3ex\hbox{$<$\kern-0.75em\raise-1.1ex\hbox{$\sim$}}\;}
\def\ga{\;  \raise0.3ex\hbox{$>$\kern-0.75em\raise-1.1ex\hbox{$\sim$}}\;}

\def\pFn{p_{\raise-0.3ex\hbox{{\scriptsize F$\!$\raise-0.03ex\hbox{\rm n}}}}
}  

\def\pFa{p_{\raise-0.3ex\hbox{{\scriptsize F$\!$\raise-0.03ex\hbox{$i$}}}}
}  

\def\pFas{p_{\raise-0.3ex\hbox{{\scriptsize F$\!$\raise-0.03ex\hbox{$k$}}}}
}  

\def\pFb{p_{\raise-0.3ex\hbox{{\scriptsize F$\!$\raise-0.03ex\hbox{$\beta$}}}}
}  

\def\vFa{v_{\raise-0.3ex\hbox{{\scriptsize F$\!$\raise-0.03ex\hbox{$i$}}}}
}  
\def\pFp{p_{\raise-0.3ex\hbox{{\scriptsize F$\!$\raise-0.03ex\hbox{\rm p}}}}
}  
\def\pFe{p_{\raise-0.3ex\hbox{{\scriptsize F$\!$\raise-0.03ex\hbox{\rm e}}}}
}  
\def\pFmu{p_{\raise-0.3ex\hbox{{\scriptsize F$\!$\raise-0.03ex\hbox{\rm
$\mu$}}}} }  
\def\m@th{\mathsurround=0pt }
\def\eqalign#1{\null\,\vcenter{\openup1\jot \m@th
   \ialign{\strut$\displaystyle{##}$&$\displaystyle{{}##}$\hfil
   \crcr#1\crcr}}\,}

\newcommand{\vp}{\mbox{${\pmb p}$}}         
\newcommand{\vps}{\mbox{${\vp '}$}}         
\newcommand{\vQ}{\mbox{${\pmb Q}$}}         
\newcommand{\vQa}{\mbox{$\vQ_{i}$}}         

\newcommand{\mN}{\mbox{$\mathcal N$}}
\newcommand{\mF}{\mbox{$\mathcal F$}}

\begin{document}

\title{Transport equations and linear response of
superfluid Fermi mixtures 
in neutron stars
} 
%
\author{Mikhail E. Gusakov}
\affiliation{
Ioffe Physical Technical Institute,
Politekhnicheskaya 26, 194021 Saint-Petersburg, Russia
}
%

\pacs{
}

\begin{abstract}
We study transport properties 
of a strongly interacting superfluid 
mixture of two Fermi-liquids.
A typical example of such matter 
is the neutron-proton liquid 
in the cores of neutron stars.
To describe the mixture, we employ the Landau theory of Fermi-liquids,
generalized to allow for the effects of superfluidity.
We formulate the kinetic equation 
and analyze linear response 
of the system to vector (e.g., electromagnetic) perturbation.
In particular, we calculate the transverse and longitudinal polarization
functions for both liquid components. 
We demonstrate, that they can be 
expressed through the Landau parameters of the mixture 
and polarization functions of noninteracting matter 
(when the Landau quasiparticle interaction is neglected). 
Our results can be used, e.g.,
for studies of the kinetic coefficients
and low-frequency long-wavelength 
collective modes in superfluid Fermi-mixtures. 
\end{abstract}

\maketitle

\section{Introduction}
\label{1}

Neutron stars are ultracompact objects 
with radius $R \sim 10$ km and a mass $л \sim M_{\odot}$,
where $M_{\odot} \approx 1.99 \times 10^{33}$ g is the solar mass.
The density in the cores of neutron stars 
is a few times larger than the density 
$\rho_0 \approx 2.8 \times 10^{14}$ g cm$^{-3}$ in atomic nuclei.
The simplest matter composition in the 
internal layers of neutron stars includes 
neutrons, protons, and electrons.
Neutrons and protons form a strongly nonideal, 
degenerate Fermi-liquid,
while electrons can be considered as a degenerate gas 
of free relativistic particles.
It is generally accepted \cite{yls99,ls01,yp04} 
that at low enough temperatures
$T \la (10^8-10^{10})$ K neutrons and protons become superfluid.
Thus, to compare the theory with observations
of neutron stars one 
should be able
to describe a strongly nonideal, 
superfluid mixture of Fermi-liquids.
In particular, 
the kinetic properties of such matter
are of primary interest.

One of the important problems in the kinetics of superfluid mixtures
is the calculation of the so called response functions
that characterize response
of the system to small external perturbations.
The response functions determine, for example, 
the emissivity of various neutrino processes 
(e.g., the Cooper-pairing neutrino emission processes 
\cite{kr04,sms07,kv08,leinson08,sr09,leinson09}
or bremsstrahlung processes \cite{lp01})
and thus influence the thermal evolution of neutron stars.
Furthermore, the formalism of response functions 
is ideally suited to study 
collective modes in the system
which may lead to a discovery 
of new exotic processes of neutrino emission.
Finally, the vector response functions determine
the screening properties of particle interaction,
influence particle collision amplitudes
and hence affect kinetic coefficients.

Although for {\it normal} matter of neutron stars
the response functions 
and the collective degrees of freedom 
have been
extensively discussed in the literature
(see, e.g., Refs.\ \cite{haensel78,md94,fm96,bs98,rplp99,gctm03,slgz03,pbsma06,lps08,dmc08,bd09a,bd09b}),
the author is unaware of any discussion 
concerning the response in {\it superfluid} 
strongly interacting mixtures.
However, for a {\it one-component} 
strongly nonideal superfluid Fermi-liquid
(neutrons in the crust of a neutron star),
the response functions were calculated recently 
by Leinson \cite{leinson09}. 
He generalized the results of Leggett \cite{leggett66}
to the case of not too small 
wave vectors and perturbation frequencies
and calculated the response 
of the system to axial-vector perturbation.

The same problem was also examined in Ref.\ \cite{kv08}.
The authors of this reference considered 
a superfluid neutron-proton mixture;
however, they do not allow for 
interaction between neutrons and protons. 
Therefore, {\it de facto} they analyze the system of
two noninteracting one-component Fermi-liquids.
The results obtained in Ref.\ \cite{kv08} 
were discussed (and criticized) in Ref.\ \cite{leinson09}.

It is convenient to calculate the response functions using the 
method of kinetic equation.
For that one has to formulate the kinetic equation
which correctly describes
(a) possible superfluidity of neutrons and protons
and 
(b) various Fermi-liquid effects connected with interactions
between neutrons and protons.
For a one-component Fermi-liquid such an equation 
was derived, for example, 
by Betbeder-Matibet and Nozieres \cite{bn69} 
(see also Ref.\ \cite{wolfle76}).
Neglecting the Fermi-liquid effects 
the superfluid kinetic equation
was thoroughly examined, e.g., 
in Refs.\ \cite{stephen65,gk69,aggk81,aggk86,kopnin01} 
(see also references therein).

There exist a number of papers
exploring the
transport properties of superfluid matter 
in the cores of neutron stars
(see, e.g., Refs.\ \cite{gy95,bhy01,sy07,sy08,acprs09}).
However, to our best knowledge, 
the kinetic equation, 
satisfying both conditions (a) and (b),
has not yet been proposed.

Partially, 
this is 
because most studies 
in the literature
have been devoted
to kinetics of electron gas, 
in particular to electron kinetic coefficients.
To calculate these coefficients the kinetic equations
for nucleons are not required  
since one usually neglects
the interaction between neutrons and electrons
and considers protons only as scatterers.
Yet another reason
is 
the one-component Fermi-liquid
(for example, electrons in metals or liquid helium-3),
for which 
the Fermi-liquid effects 
are well known to play no major role 
in a 
variety
of kinetic phenomena.

Here we demonstrate that
the situation with 
the superfluid {\it mixtures} is different
and that
the Fermi-liquid effects
can be more pronounced there.
To a large extent this is due to the mean-field interaction 
between neutron and proton quasiparticles.
In the hydrodynamics of superfluid mixtures 
\cite{ab75,bjk96,gh05,ch06,gusakov07,ch08} 
this interaction leads to 
{\it entrainment} 
of neutrons by the superfluid motion of protons 
and vice versa (the so called entrainment effect).
We will show that the similar effects are 
important 
for 
the kinetics of superfluid mixtures.

In the present paper we formulate 
the collisionless kinetic equation,
satisfying both conditions (a) and (b).
Employing this equation, 
we calculate and analyze
response of the system to a vector (electromagnetic)
perturbation
(more precisely, we calculate
the longitudinal 
and transverse polarization functions).
To describe the superfluid mixture we use 
the framework of the Landau theory of Fermi-liquids,
extended 
by Larkin and Migdal \cite{lm63,larkin64}
and by Leggett \cite{leggett65,leggett66}
to allow for superfluidity.
For simplicity, we assume that 
in thermodynamic equilibrium 
both particle species pair 
in the spin-singlet $^1$S$_0$ state 
(this is indeed a simplification since, 
according to microscopic calculations \cite{ls01}, 
neutrons in the cores of neutron stars pair
in the spin-triplet $^3$P$_2$ state).
%
%

In the collisionless approximation, 
which we are interested in here,
the electrons interact with the nucleons only through
a self-consistent electromagnetic field.
Thus, in what follows, the electrons can be safely ignored; 
this does not change our results.

The paper is organized as follows.
In Sec.\ II we derive the kinetic equation
describing, in the linear approximation,
mixtures of superfluid Fermi-liquids.
In Sec.\ III we calculate and analyze 
the longitudinal and transverse 
polarization functions for mixtures.
In Sec.\ IV we propose a nonlinear kinetic equation
describing superfluid mixtures in the quasiclassical limit.
Section V presents the summary.

Below, unless otherwise stated, 
we use the system of units in which 
the Planck constant $\hbar$,
the speed of light $c$,
the Boltzmann constant $k_{\rm B}$, 
and the normalization volume $V$
equal unity, $\hbar=c=k_{\rm B}=V=1$.

\section{The kinetic equation 
for superfluid Fermi mixtures}
\label{2}
\subsection{Thermodynamic equilibrium}

To establish notations, let us briefly consider 
a 
strongly interacting degenerate homogeneous Fermi-liquid 
in thermodynamic equilibrium at some temperature $T$.
We assume that the liquid is two-component and composed 
of particles of two species, $i=1$ and $i=2$.
Here and below the indices $i$ and $j$ 
refer to these species.

Weakly excited states of our system
can be described 
in terms of quasiparticles.
According to the Landau theory of Fermi-liquids, 
in the normal (nonsuperfluid) matter, 
the energy $\varepsilon_{\pmb k}^{(i)}$ 
of a quasiparticle $i$ with 
momentum ${\pmb k}$ 
is a functional of quasiparticle distribution functions 
$\mN_{{\pmb k}}^{(j=1, \, 2)}$ (see, e.g., Refs.\ \cite{pn66,bp91}),
\begin{equation}
\varepsilon_{\pmb k}^{(i)}\left[\mN_{\pmb k}^{(j)} \right]
=\varepsilon^{(i)}_{{\pmb k} \, 0} 
+ \sum_{{\pmb k}' \sigma' j} f^{ij}({\pmb k}, {\pmb k}') 
\left[ 
 \mN_{{\pmb k}'}^{(
 j)} 
 -\theta_{{\pmb k}'}^{(j)}
\right].
\label{energy}
\end{equation}
For nonsuperfluid matter in thermodynamic equilibrium
the distribution function $\mN_{{\pmb k}}^{(i)}$ 
is denoted $n_{{\pmb k} \, 0}^{(i)}$ and is given by
\begin{equation}
n_{{\pmb k} \, 0}^{(i)} = \left\langle a_{{\pmb k} \sigma}^{(i) \dagger} 
a_{{\pmb k} \sigma}^{(i)} \right\rangle
=\frac{1}{1 + 
{\rm e}^{\left(\varepsilon_{\pmb k}^{(i)}
\left[ n_{{\pmb k} \, 0}^{(j)}\right]-\mu_i \right)/T} },
\label{nk}
\end{equation}
In Eqs.\ (\ref{energy}) and (\ref{nk}) 
$\varepsilon_{{\pmb k} \, 0}^{(i)}$ 
is the quasiparticle energy at $T=0$; 
$\sigma$ and $\sigma' = \pm 1$ are spin indices;
$\theta_{{\pmb k}}^{(i)}=\theta(k_{{\rm F}i}-k)$, 
where $\theta(x)$ is the step function, 
and $k_{{\rm F}i}$ is the Fermi momentum.
Furthermore, 
$a_{{\pmb k} \sigma}^{(i)} =
a_{{\pmb k} \uparrow}^{(i)}$ or $a_{{\pmb k} \downarrow}^{(i)}$
is the annihilation operator of a quasiparticle
in a state (${\pmb k} \sigma$) and
$\mu_i$ is the chemical potential.
Finally, $f^{ij}({\pmb k}, {\pmb k}')$ in Eq.\ (\ref{energy})
is the 
Landau quasiparticle interaction.
In the present paper we only deal 
with the spin-unpolarized matter
and do not consider forces that rotate spin.
This allows us to disregard the spin dependence of the interaction function,
assuming that  $f^{ij}({\pmb k}, {\pmb k}')$ 
is already averaged over spin variables.
We also suppress spin indices 
in formulas, whenever possible.
Notice, however, that the kinetic equation (\ref{kineq}),
obtained below,
can be used, with minor modifications, 
to study transport properties of {\it spin-polarized} matter
[then, of course, one should take into account the spin-dependence
of $f^{ij}({\pmb k}, {\pmb k}')$].

In the vicinity of the Fermi surface the 
lengths of the vectors ${\pmb k}$ and ${\pmb k}'$ 
in the arguments of
the function $f^{i j}({\pmb k}, {\pmb k}')$
can be approximately set equal to $k \approx k_{{\rm F} i}$ 
and $k' \approx k_{{\rm F}j}$, 
while the function itself can be 
expanded into Legendre polynomials $P_l({\rm cos \, \theta})$,
\begin{equation}
f^{i j}({\pmb k}, {\pmb k}') = \sum_l f^{i j}_l \, P_l(\cos \theta),
\label{fij}
\end{equation}
where $\theta$ is the angle between ${\pmb k}$ and ${\pmb k}'$,
and $f^{i j}_l$ are the symmetric Landau parameters,
$f_l^{ij}=f_l^{ji}$.

Assume now that both particles, $i=1$ and $i=2$, are superfluid 
(and we are still in thermodynamic equilibrium at a temperature $T$).
The spin $1/2$ elementary excitations 
in superfluid Fermi-liquid are the {\it Bogoliubov} excitations.
In the absence of superfluid currents in the system, 
their energy $E_{\pmb k}^{(i)}$ and the distribution function 
$\mF_{{\pmb k} \, 0}^{(i)}$ are (see, e.g., Ref.\ \cite{lp80})
\begin{eqnarray}
E_{\pmb k}^{(i)} &=& \sqrt{\xi_{\pmb k}^{(i) 2} + \Delta_i^2},
\label{Ek}\\
\mF_{{\pmb k} \, 0}^{(i)} &=& 
\frac{1}{1+{\rm e}^{E_{\pmb k}^{(i)}/T  }},
\label{fk}
\end{eqnarray}
where $\xi_{\pmb k}^{(i)}$ equals
(see, e.g., Ref.\ \cite{gh05})
\begin{equation}
\xi_{\pmb k}^{(i)} =
\varepsilon_{\pmb k}^{(i)}\left[ \mN_{{\pmb k} \, 0}^{(j)} \right] - \mu_i =
v_{{\rm F} i} ( k-k_{{\rm F}i}) 
+ O\left[ \left(\frac{T}{\mu_j} \right)^2 + \left(\frac{\Delta_i}{\mu_j} \right)^2 \right]
\approx v_{{\rm F} i} ( k-k_{{\rm F}i}).
\label{xik}
\end{equation}
The function
$\mN_{{\pmb k} \, 0}^{(i)}$ 
will be defined in Eq.\ (\ref{nks0});
$v_{{\rm F} i}=k_{{\rm F}i}/m_i^{\ast}$ 
is the Fermi velocity 
and $m_i^{\ast}$ is the effective mass.  
In the nonrelativistic limit $m_i^{\ast}$ 
can be found from the equation \cite{bjk96,gh05,gkh09a}
\begin{equation}
\frac{m_i}{m_i^{\ast}} = 1 - \sum_j \frac{m_j G_{ij}}{n_i},
\label{effmass}
\end{equation}
where 
\begin{equation}
n_i=\frac{p_{{\rm F}i}^3}{3 \pi^2}
\label{ni}
\end{equation}
is the number density;
$m_i$ is the bare mass;
and the symmetric matrix $G_{ij}$ is defined as
\begin{equation}
G_{ij} \equiv \frac{1}{9 \pi^4} \, p_{{\rm F}i}^2 p_{{\rm F}j}^2 \, f_1^{ij}.
\label{Gij}
\end{equation}

Furthermore, $\Delta_i$ in Eqs.\ (\ref{Ek})--(\ref{xik}) 
is the energy gap in thermodynamic equilibrium. 
Because we assume the singlet-state 
$^1{\rm S}_0$ pairing of quasiparticles,
the gap $\Delta_i$ depends only on 
$k=|{\pmb k}|$.
It can be found from the standard equation
\begin{equation}
\Delta_i(k) = 
-\sum_{{\pmb k}'} V^{(i)}({\pmb k}, {\pmb k}') 
\,\, \Delta_{i}(k') \,\, F^{(i)}_{{\pmb k}'}.
\label{gap}
\end{equation}
Here we define 
\begin{equation}
F^{(i)}_{\pmb k} \equiv  \frac{1}{2 E_{\pmb k}^{(i)}} \, 
{\rm tanh}\left( \frac{E_{\pmb k}^{(i)}}{2 T} \right).
\label{Fk}
\end{equation}
In Eq.\ (\ref{gap}) $V^{(i)}({\pmb k}, {\pmb k}')$ 
is the pairing potential for particles $i$.
In analogy with the function $f^{ij}({\pmb k}, {\pmb k}')$ [see Eq.\ (\ref{fij})],
we expand it
into Legendre polynomials,
\begin{equation}
V^{(i)}({\pmb k},{\pmb k}') = \sum_l V_l^{(i)} \,\, P_{l}(\cos \theta).
\label{Vkk}
\end{equation}

Near the Fermi surface a smoothly varying function $\Delta_i(k)$
can be approximated as $\Delta_i(k) \approx \Delta_i(k_{{\rm F}i})$ 
(see, e.g., Ref.\ \cite{bn69}). 
Then, combining Eqs.\ (\ref{gap}) and (\ref{Vkk}), 
one obtains the following equation for $\Delta_i$, 
\begin{equation}
1 = - V_0^{(i)} \, \sum_{{\pmb k}'} \, F^{(i)}_{{\pmb k}'}.
\label{Delta}
\end{equation}

As follows from Eq.\ (\ref{fk}), 
the distribution function $\mathcal{F}_{{\pmb k} \, 0}^{(i)}$
for {\it Bogoliubov} excitations is a scalar quantity.
In contrast, the distribution function 
${\pmb {\rm n}}_{{\pmb k}\sigma \, 0}^{(i)}$
for {\it Landau} quasiparticles in superfluid matter 
is a matrix rather than a scalar. 
In thermodynamic equilibrium it can be written as 
(see, e.g., Ref.\ \cite{bn69})
\begin{equation}
{\pmb {\rm n}}_{{\pmb k}\sigma \, 0}^{(i)} =
\left(
\begin{array}{ccc}
\left\langle a_{{\pmb k} \sigma}^{(i) \dagger} a_{{\pmb k} \sigma}^{(i)} \right\rangle & 
 \left\langle a_{{\pmb k} \sigma}^{(i) \dagger} a_{-{\pmb k} -\sigma}^{(i) \dagger} \right\rangle\\
\left\langle a_{-{\pmb k} -\sigma}^{(i)} a_{{\pmb k} \sigma}^{(i)} \right\rangle & 
\left\langle a_{-{\pmb k} -\sigma}^{(i)} a_{-{\pmb k} -\sigma}^{(i) \dagger} \right\rangle
\end{array}
\right) = \frac{1}{2} \left( \hat{{\pmb {\rm 1}}} 
- 2\,\, {\pmb \epsilon}_{{\pmb k} \sigma \, 0}^{(i)} \, F^{(i)}_{\pmb k} \right),
\label{nks00}
\end{equation}
where $\hat{{\pmb 1}}$ is the unit matrix.
The `energy matrix' ${\pmb \epsilon}_{{\pmb k} \sigma \, 0}^{(i)}$ 
in Eq.\ (\ref{nks00}) equals
\begin{equation}
{\pmb \epsilon}_{{\pmb k} \sigma \, 0}^{(i)} =
\left(
\begin{array}{ccc}
\xi_{\pmb k}^{(i)} & 
 \sigma \Delta_i \\
\sigma \Delta_i & 
-\xi_{\pmb k}^{(i)}
\end{array}
\right).
\label{eks0}
\end{equation}
For {\it superfluid} matter 
in the {\it absence} of superfluid currents
the average {\it equilibrium} number $\mN_{{\pmb k} \, 0}^{(i)}$
of Landau quasiparticles in a state $({\pmb k}\sigma)$
is given by the element 
$\left\langle a_{{\pmb k} \sigma}^{(i) \dagger} a_{{\pmb k} \sigma}^{(i)} \right\rangle$
of the matrix ${\pmb {\rm n}}_{{\pmb k}\sigma \, 0}^{(i)}$,
\begin{equation}
\mN_{{\pmb k} \, 0}^{(i)} \equiv 
\left\langle a_{{\pmb k} \sigma}^{(i) \dagger} a_{{\pmb k} \sigma}^{(i)} \right\rangle =
\frac{1}{2} 
\left( 1 - 2\,\, \xi_{\pmb k}^{(i)} \, F^{(i)}_{\pmb k} \right)
\label{nks0}
\end{equation}
[compare this expression with the corresponding 
Eq.~(\ref{nk}) for normal matter].

\subsection{The system of kinetic equations}

To obtain the kinetic equation let us slightly perturb the system. 
Since our aim in this section is to determine the kinetic equation 
in the {\it linear approximation}, 
we may assume, 
without any loss of generality,
that the perturbation 
varies with coordinate ${\pmb r}$ and time $t$ as
${\rm e}^{i ({\pmb q}{\pmb r}-\omega t)}$, 
where ${\pmb q}$ and $\omega$ 
are the perturbation wave vector and frequency, respectively.
To use the Landau theory of Fermi-liquids
we have to assume, in addition, that
$q\ll k_{{\rm F}i}$ and $\omega \ll \mu_i$.

The only non-zero matrix elements, induced by the perturbation,
can be written in a compact form as (see, e.g., \cite {bn69})
\begin{equation}
\delta {\pmb {\rm n}}_{{\pmb k}\sigma}^{(i)}({\pmb q})=
\left(
\begin{array}{ccc}
\left\langle a_{{\pmb k}_- \sigma}^{(i) \dagger} a_{{\pmb k}_+ \sigma}^{(i)} \right\rangle & 
 \left\langle a_{{\pmb k}_- \sigma}^{(i) \dagger} a_{-{\pmb k}_+ -\sigma}^{(i) \dagger} \right\rangle\\
\left\langle a_{-{\pmb k}_- -\sigma}^{(i)} a_{{\pmb k}_+ \sigma}^{(i)} \right\rangle & 
\left\langle a_{-{\pmb k}_- -\sigma}^{(i)} a_{-{\pmb k}_+ -\sigma}^{(i) \dagger} \right\rangle
\end{array}
\right)
\equiv \left(
\begin{array}{ccc}
\delta n_{{\pmb k}\sigma \, 11}^{(i)} & 
 \delta n_{{\pmb k}\sigma \, 12}^{(i)}\\
  \delta n_{{\pmb k}\sigma \, 21}^{(i)}& 
  \delta n_{{\pmb k}\sigma \, 22}^{(i)}
\end{array}
\right).
\label{delta_n}
\end{equation}
Here and below we use the notation
\begin{equation}
{\pmb k}_-={\pmb k} - \frac{{\pmb q}}{2}, \quad \quad 
{\pmb k}_+={\pmb k} + \frac{{\pmb q}}{2}.
\label{k1k2}
\end{equation}
The matrix $\delta {\pmb {\rm n}}_{{\pmb k}\sigma}^{(i)}({\pmb q})$ 
with the elements $\delta n_{{\pmb k}\sigma \, 11}^{(i)},
\ldots,\delta n_{{\pmb k}\sigma \, 22}^{(i)}$,
defined in Eq.\ (\ref{delta_n}), can be interpreted as 
a small deviation from the equilibrium distribution function 
${\pmb {\rm n}}_{{\pmb k}\sigma \, 0}^{(i)}$, 
caused by the perturbation. 
The collisionless kinetic equation 
for $\delta {\pmb {\rm n}}_{{\pmb k}\sigma}^{(i)}({\pmb q})$
can be found following the derivation
of Ref.\ \cite{bn69}.
The result is:
\begin{equation}
\omega \, \delta {\pmb {\rm n}}_{{\pmb k}\sigma}^{(i)}
= \delta {\pmb {\rm n}}_{{\pmb k}\sigma}^{(i)} \,\,
{\pmb \epsilon}_{{\pmb k}_+ \sigma \, 0}^{(i)}
-{\pmb \epsilon}_{{\pmb k}_- \sigma \, 0}^{(i)} \,\,
\delta {\pmb {\rm n}}_{{\pmb k}\sigma}^{(i)}
+ {\pmb {\rm n}}_{{\pmb k}_- \sigma \, 0}^{(i)} \,\,
\delta {\pmb \epsilon}_{{\pmb k}\sigma}^{(i)} 
- \delta {\pmb \epsilon}_{{\pmb k}\sigma}^{(i)} \,\,
{\pmb {\rm n}}_{{\pmb k}_+ \sigma \, 0}^{(i)}.
\label{kineq}
\end{equation}
In this equation 
$\delta {\pmb \epsilon}_{{\pmb k}\sigma}^{(i)}$ is a matrix,
describing local deviation of the quasiparticle energy 
from its equilibrium value ${\pmb \epsilon}_{{\pmb k} \sigma \, 0}^{(i)}$. 
It is the sum of two terms,
\begin{equation}
\delta {\pmb \epsilon}_{{\pmb k}\sigma}^{(i)} 
= \lambda_{{\pmb k}\sigma}^{(i)} + \Lambda^{(i)}_{{\pmb k}\sigma},
\label{dek}
\end{equation}
where the term $\lambda_{{\pmb k}\sigma}^{(i)}$ 
describes the change 
of the quasiparticle energy with the distribution function,
\begin{equation}
\lambda_{{\pmb k} \sigma}^{(i)} = 
\left(
\begin{array}{ccc}
\sum_{{\pmb k}' \sigma' j} \,\, f^{ij}({\pmb k}, {\pmb k}') 
\,\, \delta n_{{\pmb k}' \sigma' \, 11}^{(j)} \quad \quad &
\sum_{{\pmb k}'} \,\,
V^{(i)}({\pmb k}, {\pmb k}') \,\, \delta n_{{\pmb k}' \sigma \, 12}^{(i)} \\
\sum_{{\pmb k}'} \,\,
V^{(i)}({\pmb k}, {\pmb k}') \,\, \delta n_{{\pmb k}' \sigma \, 21}^{(i)} \quad \quad &
\sum_{{\pmb k}' \sigma' j} \,\, f^{ij}({\pmb k}, {\pmb k}') 
\,\, \delta n_{{\pmb k}' \sigma' \, 22}^{(j)} 
\end{array}
\right),
\label{lambda}
\end{equation}
and the term $\Lambda^{(i)}_{{\pmb k}\sigma}$
is responsible
for the interaction of quasiparticles with 
the {\it self-consistent} electromagnetic field,
\begin{equation}
\Lambda^{(i)}_{{\pmb k}\sigma} = 
\left(
\begin{array}{ccc}
e_i \, V - \alpha_i \,\, \frac{{\pmb k} {\pmb A}}{m_i}& 
0\\
0& 
- e_i \, V - \alpha_i \,\, \frac{{\pmb k} {\pmb A}}{m_i}
\end{array}
\right).
\label{Lambda_ext}
\end{equation}
Here $V$ and ${\pmb A}$ are 
the scalar and vector electromagnetic potentials, respectively.
It is assumed that 
in the unperturbed system
$V=0$ and ${\pmb A}=0$.

Equations (\ref{kineq})--(\ref{Lambda_ext}) are trivial 
generalizations, to the case of superfluid mixtures, 
of the corresponding equations presented in Ref.\ \cite{bn69}.
The only non-trivial point is the expression for the coefficient $\alpha_i$.
In Ref.\ \cite{bn69} $\alpha_i$ is equal to 
the quasiparticle electric charge $e_i$.
This result is
valid only for a one-component Fermi-liquid.
As is demonstrated in the Appendix A,
for a mixture of Fermi-liquids one should instead write
\begin{equation}
\alpha_i = \frac{m_i}{n_i} \sum_j e_j \, Y_{ij},
\label{alpha_i}
\end{equation}
where $Y_{ij}$ is the relativistic entrainment matrix 
at zero temperature, given by \cite{gkh09a,gkh09b}
\begin{equation}
Y_{ij} = \frac{n_i}{m_i^{\ast}} \, \delta_{ij} + G_{ij}.
\label{Yik2}
\end{equation}
In this equation $\delta_{ij}$ is the Kronecker symbol 
and the matrix $G_{ij}$ is defined 
in Eq.\ (\ref{Gij}).

The kinetic equation ({\ref{kineq}}) consists 
of four coupled integral equations.
Their solution determines 
the matrix $\delta {\pmb {\rm n}}_{{\pmb k}\sigma}^{(i)}({\pmb q})$
[i.e., the four functions $\delta n_{{\pmb k}\sigma \, 11}^{(i)},
\ldots,\delta n_{{\pmb k}\sigma \, 22}^{(i)}$].

The system (\ref{kineq}) 
can be substantially simplified by introducing 
a new set of variables
\begin{eqnarray}
\delta n_{{\pmb k} \sigma +}^{(i)} &=& \delta n_{{\pmb k} \sigma \, 11}^{(i)} 
+ \delta n_{{\pmb k} \sigma \, 22}^{(i)},
\label{dnk+}\\
\delta n_{{\pmb k} \sigma -}^{(i)} &=& \delta n_{{\pmb k} \sigma \, 11}^{(i)} 
- \delta n_{{\pmb k} \sigma \, 22}^{(i)},
\label{dnk-}\\
\delta s_{{\pmb k} \sigma +}^{(i)} &=& \sigma  \left[\delta n_{{\pmb k} \sigma \, 12}^{(i)} 
+ \delta n_{{\pmb k} \sigma \, 21}^{(i)} \right],
\label{dsk+}\\
\delta s_{{\pmb k} \sigma -}^{(i)} &=& \sigma  \left[\delta n_{{\pmb k} \sigma \, 12}^{(i)} 
- \delta n_{{\pmb k} \sigma \, 21}^{(i)}\right],
\label{dsk-}
\end{eqnarray}
with the obvious symmetry properties [see Eq.\ (\ref{delta_n})],
\begin{eqnarray}
\delta n_{{\pmb k} \sigma +}^{(i)} &=& -\delta n_{-{\pmb k} -\sigma \, +}^{(i)},
\label{dnk+sym}\\
\delta n_{{\pmb k} \sigma -}^{(i)} &=& \delta n_{-{\pmb k} -\sigma \, -}^{(i)}, 
\label{dnk-sym}\\
\delta s_{{\pmb k} \sigma +}^{(i)} &=& \delta s_{-{\pmb k} -\sigma \, +}^{(i)}, 
\label{dsk+sym}\\
\delta s_{{\pmb k} \sigma -}^{(i)} &=& \delta s_{-{\pmb k} -\sigma \, -}^{(i)}.
\label{dsk-sym}
\end{eqnarray}
Using these variables, the system of kinetic equations (\ref{kineq}) can be
rewritten in the form
\begin{eqnarray}
\omega \,\, \delta n_{{\pmb k} \sigma +}^{(i)} &=&
\left[F_{{\pmb k}_+}^{(i)} \xi_{{\pmb k}_+}^{(i)} 
- F_{{\pmb k}_-}^{(i)} \xi_{{\pmb k}_-}^{(i)}\right]
\, V_{{\pmb k}\sigma}^{(i)} 
+ \left[ \xi_{{\pmb k}_+}^{(i)}  -\xi_{{\pmb k}_-}^{(i)}  \right] \,
\delta n_{{\pmb k} \sigma -}^{(i)} 
\nonumber\\
&+& \Delta_i \, \left[F_{{\pmb k}_+}^{(i)} - F_{{\pmb k}_-}^{(i)} \right] 
\,  O_{{\pmb k} \sigma +}^{(i)},
\label{kineq1}\\
\omega \,\, \delta n_{{\pmb k} \sigma -}^{(i)} &=& 
\left[F_{{\pmb k}_-}^{(i)} \xi_{{\pmb k}_-}^{(i)} 
- F_{{\pmb k}_+}^{(i)} \xi_{{\pmb k}_+}^{(i)}\right] 
\, A_{{\pmb k}\sigma}^{(i)} 
+ \left[ \xi_{{\pmb k}_+}^{(i)}  -\xi_{{\pmb k}_-}^{(i)}  \right] 
\, \delta n_{{\pmb k} \sigma +}^{(i)} 
\nonumber\\
&+& 2 \,\,\Delta_i \, \, \delta s_{{\pmb k} \sigma -}^{(i)} 
+ \Delta_i \, 
\left[F_{{\pmb k}_-}^{(i)} + F_{{\pmb k}_+}^{(i)} \right] 
\, O_{{\pmb k} \sigma -}^{(i)},
\label{kineq2}\\
\omega \,\, \delta s_{{\pmb k} \sigma -}^{(i)} &=& 
\Delta_i \, \left[F_{{\pmb k}_-}^{(i)} + F_{{\pmb k}_+}^{(i)} \right] 
\, V_{{\pmb k}\sigma}^{(i)} 
+ 2 \,\, \Delta_i \, \, \delta n_{{\pmb k} \sigma -}^{(i)} 
\nonumber\\
&-& \left[ \xi_{{\pmb k}_-}^{(i)}  + \xi_{{\pmb k}_+}^{(i)}  \right]
\, \delta s_{{\pmb k} \sigma +}^{(i)} 
- \left[F_{{\pmb k}_-}^{(i)} \xi_{{\pmb k}_-}^{(i)} 
+ F_{{\pmb k}_+}^{(i)} \xi_{{\pmb k}_+}^{(i)}\right] \, O_{{\pmb k} \sigma +}^{(i)},
\label{kineq3}\\
\omega \,\, \delta s_{{\pmb k} \sigma +}^{(i)} &=& 
\Delta_i \, \left[F_{{\pmb k}_-}^{(i)} - F_{{\pmb k}_+}^{(i)} \right] 
 \, A_{{\pmb k}\sigma}^{(i)} 
 \nonumber\\
 &-& \left[ \xi_{{\pmb k}_-}^{(i)}  + \xi_{{\pmb k}_+}^{(i)}  \right] \,
 \delta s_{{\pmb k} \sigma -}^{(i)}
 - \left[F_{{\pmb k}_-}^{(i)} \xi_{{\pmb k}_-}^{(i)} 
+ F_{{\pmb k}_+}^{(i)} \xi_{{\pmb k}_+}^{(i)}\right] \, O_{{\pmb k} \sigma -}^{(i)},
\label{kineq4}
\end{eqnarray}
where $O_{{\pmb k} \sigma +}^{(i)}$ 
and $O_{{\pmb k} \sigma -}^{(i)}$
equal
\begin{eqnarray}
O_{{\pmb k} \sigma +}^{(i)} &=& \sum_{{\pmb k}'} 
V^{(i)}({\pmb k}, {\pmb k}') \, \delta s_{{\pmb k}' \sigma +}^{(i)},
\label{O+}\\
O_{{\pmb k} \sigma -}^{(i)} &=& \sum_{{\pmb k}'} 
V^{(i)}({\pmb k}, {\pmb k}') \, \delta s_{{\pmb k}' \sigma -}^{(i)},
\label{O-}
\end{eqnarray}
and the functions $V_{{\pmb k}\sigma}^{(i)}$ and $A_{{\pmb k}\sigma}^{(i)}$
are defined as
\begin{eqnarray}
V_{{\pmb k} \sigma}^{(i)} &=& 2 e_i V+\sum_{{\pmb k}' \sigma' j} 
f^{ij}({\pmb k}, {\pmb k}') \, \delta n_{{\pmb k}' \sigma' -}^{(j)},
\label{Vks}\\
A_{{\pmb k} \sigma}^{(i)} &=& 
2 \alpha_i \,\, \frac{{\pmb k} {\pmb A}}{m_i} 
-\sum_{{\pmb k}' \sigma' j} 
f^{ij}({\pmb k}, {\pmb k}') \, \delta n_{{\pmb k}' \sigma' +}^{(j)}.
\label{Aks}
\end{eqnarray}
The system (\ref{kineq1})--(\ref{kineq4}) 
contains all information to calculate 
linear gauge-invariant response 
of the two-component superfluid Fermi-liquid 
to a vector (e.g., electromagnetic) perturbation.

\subsection{The particle current density}

Let us sum Eq.\ (\ref{kineq2}) over ${\pmb k}$ and $\sigma$.
Then, using Eq.\ (\ref{Delta}) one obtains
that, with the accuracy up to quadratic terms in $q/k_{{\rm F}i} \ll 1$, 
the last two terms in the right-hand side of Eq.\ (\ref{kineq2}) 
cancel out and we are left with
\begin{equation}
\omega \, \sum_{{\pmb k}\sigma} \delta n_{{\pmb k} \sigma -}^{(i)} =
\sum_{{\pmb k}\sigma} 
\left[ \mN_{{\pmb k}_+ \, 0}^{(i)}-\mN_{{\pmb k}_- \, 0}^{(i)} \right]
 A_{{\pmb k}\sigma}^{(i)} 
+ \sum_{{\pmb k}\sigma} 
\left[ \xi_{{\pmb k}_+}^{(i)}  -\xi_{{\pmb k}_-}^{(i)}  \right] 
\, \delta n_{{\pmb k} \sigma +}^{(i)}.
\label{kineq22}
\end{equation}
Here we also made use of Eq.\ (\ref{nks0}).
To proceed further, we expand 
$\xi_{{\pmb k}_\pm}^{(i)}$ in Taylor series, 
\begin{equation}
\xi_{{\pmb k}_\pm}^{(i)}=\xi_{\pmb k}^{(i)} \pm \frac{{\pmb q} {\pmb v}_i}{2} 
+ O\left[\left(\frac{q}{k_{{\rm F}i}}\right)^2\right],
\label{kineq222}
\end{equation}
where ${\pmb v}_i$ is the velocity of Landau quasiparticles on the Fermi surface,
${\pmb v}_i \equiv v_{{\rm F}i} \, \, ({\pmb k}/k)$. 

Eq.\ (\ref{kineq22}) can be additionally simplified 
with the help of Eq.\ (\ref{Aks})
and the expansion (\ref{kineq222}).
Neglecting all terms of the second and higher orders in $q/k_{{\rm F}i}$, 
one obtains particle conservation law
\begin{equation}
\omega \, \delta n_i = {\pmb q} \, {\pmb j}_i
\label{conservation}
\end{equation}
with 
\begin{eqnarray}
\delta n_i &=& \frac{1}{2} \, \sum_{{\pmb k} \sigma} 
\, \delta n_{{\pmb k} \sigma -}^{(i)},
\label{density}\\
{\pmb j}_i &=& \frac{1}{2} \, \sum_{{\pmb k} \sigma} 
\left[ {\pmb v}_i \,\, \delta n_{{\pmb k}\sigma +}^{(i)} + 
\mN_{{\pmb k} \, 0}^{(i)} \,\, \frac{\partial}{\partial {\pmb k}} 
\sum_{{\pmb k}' \sigma' j} \, f^{ij}({\pmb k}, {\pmb k}') 
\,\, \delta n_{{\pmb k}' \sigma' +}^{(j)} \right] 
- \alpha_i \, \frac{n_i}{m_i} \, {\pmb A}.
\label{Ji}
\end{eqnarray}
Because $\mN_{{\pmb k} \, 0}^{(i)}$ 
is the isotropic equilibrium distribution function, 
one has $\partial \mN_{{\pmb k} \, 0}^{(i)}/\partial {\pmb k}=
[\partial \mN_{{\pmb k} \, 0}^{(i)}/\partial k] \, ({\pmb k}/k)$,
so that Eq.\ (\ref{Ji}) can be rewritten as
\begin{equation}
{\pmb j}_i = \frac{1}{2} \, \sum_{{\pmb k} \sigma} 
\left[ {\pmb v}_i \,\, \delta n_{{\pmb k}\sigma +}^{(i)} 
- \sum_{{\pmb k}' \sigma' j} 
\frac{\partial \mN_{{\pmb k}\, 0}^{(i)}}{\partial k}  
\,\,  f_1^{ij}\,\, \frac{{\pmb k}\, ({\pmb k}\,{\pmb k}')}{k^2 \, k'} 
\,\, \delta n_{{\pmb k}' \sigma' +}^{(j)}
\right] 
- \alpha_i \, \frac{n_i}{m_i} \, {\pmb A}.
\label{Ji2}
\end{equation}
where the second term in square brackets 
was integrated by parts and we made use of the expansion (\ref{fij}).

The quantities $\delta n_i$ and ${\pmb j}_i$ 
in Eqs.\ (\ref{conservation})--(\ref{Ji2})
can be interpreted 
as the number density perturbation
and particle current density, respectively.
To prove that this is really so,
we employ Eqs.\ (\ref{dnk+}) and (\ref{dnk-}) and the symmetry 
relations (\ref{dnk+sym}) and (\ref{dnk-sym}),
and rewrite Eqs.\ (\ref{density}) and (\ref{Ji}) 
in the familiar form,
\begin{eqnarray}
\delta n_i &=& \sum_{{\pmb k} \sigma} 
\, \delta n_{{\pmb k} \sigma \, 11}^{(i)},
\label{density1}\\
{\pmb j}_i &=& \sum_{{\pmb k} \sigma} 
\left[ {\pmb v}_i \,\, \delta n_{{\pmb k}\sigma \, 11}^{(i)} + 
\mN_{{\pmb k}\, 0}^{(i)} \,\, \frac{\partial}{\partial {\pmb k}} 
\sum_{{\pmb k}' \sigma' j} \, f^{ij}({\pmb k}, {\pmb k}') 
\,\, \delta n_{{\pmb k}' \sigma' \, 11}^{(j)} \right] 
- \alpha_i \, \frac{n_i}{m_i} \, {\pmb A}.
\label{Ji1}
\end{eqnarray}
One sees that $\delta n_i$ is indeed the number density perturbation.
Thus, as follows from the continuity equation (\ref{conservation}), 
${\pmb j}_i$ is the particle current density.
Notice that, ${\pmb j}_i$ is formally given by the same expression as for
normal (nonsuperfluid) Fermi-liquid (see, e.g., Refs.\ \cite{pn66,bp91}).
For a one-component Fermi-liquid this was first 
demonstrated by Leggett \cite{leggett65}.

Since the quantities $\delta n_i$ and ${\pmb j}_i$ 
are observables,
they {\it must be invariant} under gauge transformations.
This property of $\delta n_i$ and ${\pmb j}_i$ is very important
for the consideration below.
For a one-component Fermi-liquid 
the gauge invariance of $\delta n_i$ and ${\pmb j}_i$, 
defined in Eqs.\ (\ref{density})--(\ref{Ji1}), 
was explicitly demonstrated by Betbeder-Matibet and Nozieres 
\cite{bn69} [see their Eq.\ (28)].

\section{Landau Fermi-liquid effects and gauge invariance}

In this section we study the response of the system
to an applied harmonic electromagnetic field.
Our aim will be to express 
the exact polarization tensor for a strongly interacting 
two-component Fermi-liquid 
through that of noninteracting liquid 
[for which the Landau quasiparticle interaction 
$f^{ij}({{\pmb k}, {\pmb k}'})=0$].

\subsection{Gauge-invariant expressions 
for $\delta n_i$ and ${\pmb j}_i$} 

Assume that $V$ and ${\pmb A}$ are some {\it self-consistent} (not external)
vector and scalar electromagnetic potentials in an {\it arbitrary} gauge.
The vector potential ${\pmb A}$ can be presented in the form
\begin{equation}
{\pmb A}={\pmb A}_{\rm l}+{\pmb A}_{\rm tr},
\label{A}
\end{equation}
where ${\pmb A}_{\rm l}={\pmb q} \, ({\pmb q} {\pmb A})/q^2$ 
is the longitudinal component of ${\pmb A}$ directed along ${\pmb q}$ 
and ${\pmb A}_{\rm tr}={\pmb A}-{\pmb A}_{\rm l}$ 
is the transverse component.

Performing a gauge transformation 
\begin{eqnarray}
\widetilde{V}&=&V+ \frac{\partial \phi}{\partial t}=V-i \, \omega \, \phi,
\label{gaugeV1}\\
\widetilde{{\pmb A}}&=&{\pmb A} - {\pmb \nabla} \phi={\pmb A}-i \, {\pmb q} \,  \phi,
\label{gaugeA1}
\end{eqnarray}
one can choose the new potentials 
$\widetilde{V}$ and $\widetilde{{\pmb A}}$
in such a way that $\widetilde{{\pmb A}}_{\rm l}=0$ (i.e. $\phi = -i A_{\rm l}/q$).
In this new gauge the relation between $\delta n_i$, ${\pmb j}_i$ 
and $\widetilde{V}$, $\widetilde{{\pmb A}}$ can generally be written as
\begin{eqnarray}
\delta n_i &=& P^{(i)}_{00} \,\, \widetilde{V},
\label{dndndn}\\
{\pmb j}_i &=&  P^{(i)}_{\rm tr} \,\, \widetilde{{\pmb A}}_{\rm tr}
+\frac{\pmb q}{\omega} \,\, P^{(i)}_{\rm l}  \,\, \widetilde{V},
\label{jjj}
\end{eqnarray}
where $P^{(i)}_{\rm l}$ and $P^{(i)}_{\rm tr}$
are, respectively, 
the {\it exact} longitudinal and transverse
polarization functions for a strongly interacting Fermi-mixture.
The function $P^{(i)}_{00}$ is related to $P^{(i)}_{\rm l}$
by the continuity equation (\ref{conservation}),
\begin{equation}
 P^{(i)}_{\rm l}=\frac{\omega^2}{q^2} \,\, P^{(i)}_{00}. 
\label{PlP00}
\end{equation}
A direct (but not easy) way to obtain 
the quantities  $P^{(i)}_{00}$, $P^{(i)}_{\rm l}$, 
and $P^{(i)}_{\rm tr}$ is to solve
the system of integral kinetic equations 
(\ref{kineq1})--(\ref{kineq4})
for $\delta n_{{\pmb k}\sigma -}^{(i)}$ and 
$\delta n_{{\pmb k}\sigma +}^{(i)}$,
and then to make use of Eqs.\ (\ref{density}) and (\ref{Ji}).

Employing Eqs.\ (\ref{gaugeV1}) and (\ref{gaugeA1}),
one can rewrite Eqs.\ (\ref{dndndn}) and (\ref{jjj}) 
in the original gauge,
\begin{eqnarray}
\delta n_i &=& P^{(i)}_{00} \left(V - \frac{\omega}{q}\, A_{\rm l} \right),
\label{dn_arbitrary}\\
{\pmb j}_i &=&  P^{(i)}_{\rm tr} \,\, {\pmb A}_{\rm tr}
+\frac{\pmb q}{\omega} \,\, P^{(i)}_{\rm l}  
\left(V -\frac{\omega}{q} \,\, A_{\rm l}\right).
\label{j_arbitrary}
\end{eqnarray}
Clearly, these expressions are gauge-invariant.
The knowledge of 
$P^{(i)}_{00}$, $P^{(i)}_{\rm l}$, 
and $P^{(i)}_{\rm tr}$
allows one to determine the longitudinal $\varepsilon_{\rm l}$ 
and transverse $\varepsilon_{\rm tr}$ dielectric functions 
of the liquid,
\begin{eqnarray}
\varepsilon_{\rm l} &=& 1 - \frac{4 \pi}{\omega^2}  \,\,
\sum_i e_i \, P_{\rm l}^{(i)} 
=1 - \frac{4 \pi}{q^2}  \,\,
\sum_i e_i \,  P_{00}^{(i)},
\label{e_l}\\
\varepsilon_{\rm tr} &=& 1 - \frac{4 \pi}{\omega^2}  \,\,
\sum_i e_i \, P_{\rm tr}^{(i)}.
\label{e_tr}
\end{eqnarray}

For noninteracting Fermi-liquid, 
for which $f^{ij}({\pmb k}, {\pmb k}')=0$,
Eqs.\ (\ref{dn_arbitrary}) and (\ref{j_arbitrary}) 
take the form
\begin{eqnarray}
\delta n_i &=& \Pi^{(i)}_{00} \left(V - \frac{\omega}{q}\, A_{\rm l} \right),
\label{dn_arbitrary2}\\
{\pmb j}_i &=&  \Pi^{(i)}_{\rm tr} \,\, {\pmb A}_{\rm tr}
+\frac{\pmb q}{\omega} \,\, \Pi^{(i)}_{\rm l}  
\left(V -\frac{\omega}{q} \,\, A_{\rm l}\right),
\label{j_arbitrary2}
\end{eqnarray}
where the {\it exact} polarization functions 
of noninteracting matter are denoted as
$\Pi^{(i)}_{00}$, $\Pi^{(i)}_{\rm l}$, and $\Pi^{(i)}_{\rm tr}$.
These quantities were carefully analyzed in the literature 
(see, e.g., Refs.\ \cite{nambu60,lp81,alf06,leinson09});
they depend on a number of parameters, 
in particular, on the particle Fermi-momentum $p_{{\rm F} i}$ 
and on the mass $m_i$.

\subsection{Landau quasiparticle interaction and the polarization functions} 

Let us calculate the exact polarization functions $P^{(i)}_{00}$, 
$P^{(i)}_{\rm l}$, and $P^{(i)}_{\rm tr}$ 
under the simplified assumption that all the Landau parameters 
except for $f^{ij}_0$ and $f^{ij}_1$ vanish,
$f^{ij}_l=0$ for $l \geq 2$.
Eqs.\ (\ref{Vks}) and (\ref{Aks}) can then be rewritten as
\begin{eqnarray}
V_{{\pmb k}\sigma}^{(i)} &=& 
2  e_i V + 2 \sum_j f_0^{ij} \, \delta n_j 
\equiv 2 e_i V_{\rm eff}^{(i)},
\label{Vks1}\\
A_{{\pmb k}\sigma}^{(i)} &=& 
2 \alpha_i \,\, \frac{{\pmb k} {\pmb A}}{m_i} 
- \sum_j \frac{{\pmb k}}{k_{{\rm F}i}\, k_{{\rm F}j}} \, f_1^{ij} 
\sum_{{\pmb k}' \sigma'} {\pmb k}' \, \delta n_{{\pmb k}' \sigma' +}^{(j)} 
\equiv 2  e_i \frac{{\pmb k} {\pmb A}_{\rm eff}^{(i)}}{m_i^\ast},
\label{Aks1}
\end{eqnarray}
where we employed Eqs.\ (\ref{dnk+sym}), (\ref{dnk-sym}), (\ref{density}), 
and introduced the {\it effective} scalar $V_{\rm eff}^{(i)}$
and vector ${\pmb A}_{\rm eff}^{(i)}$ electromagnetic potentials.
One can express them in terms of 
real electromagnetic potentials $V$ and ${\pmb A}$
and {\it gauge-invariant}
observables
$\delta n_i$ and ${\pmb j}_i$
by making use of Eqs.\ (\ref{density}) and (\ref{Ji}), respectively.
The result is:
\begin{eqnarray}
V_{\rm eff}^{(i)} &=& 
V + \frac{1}{e_i}\, \sum_j f_0^{ij} \, \delta n_j,
\label{Veff1}\\
{\pmb A}_{\rm eff}^{(i)} &=& 
{\pmb A} + \frac{1}{e_i} \, \sum_j \gamma_{ij} \, {\pmb j}_j.
\label{Aeff1}
\end{eqnarray}
To obtain Eq.\ (\ref{Aeff1}) we use 
Eq.\ (\ref{effmass}) for the effective mass $m_i^\ast$
and the expression (\ref{alpha_i}) for $\alpha_i$.
The detailed derivation of Eq.\ (64) for a one-component Fermi-liquid 
is given in Appendix B.
%
The matrix $\gamma_{ij}$ in Eq.\ (\ref{Aeff1}) 
depends on the Landau parameters $f_1^{ij}$
and equals
\begin{eqnarray} 
\gamma_{ii} &=& \frac{m_i^2}{S_i} \, 
\left( G_{i i} G_{i j} \, m_{i } 
+ G_{i j}^2  \, m_{j} - G_{i i} \, n_j \right),
\label{gammaii}\\
\gamma_{i j} &=& \frac{m_{i} m_{j} }{S_i} \, \,
G_{i j} \left(  G_{ ii} \, m_i  
+ G_{i j} \, m_{j}   -n_i      \right),
\label{gammaij}
\end{eqnarray}
with
\begin{equation}
S_{i} = \left(   G_{i i} \, m_{i} 
+ G_{i j} \, m_{j} - n_{i}  \right) 
\left(   G_{i j} \, m_{i} \, n_{i} 
+ G_{i j } \, m_{j} \, n_{j} - n_{i} \, n_{j}  \right).
\label{Si}
\end{equation}
Here the symmetric matrix $G_{ij}$ is defined by Eq.\ (\ref{Gij}). 
In Eqs.\ (\ref{gammaii})--(\ref{Si}) 
the indices $i$ and $j$ belong to different 
particle species, $i \neq j$.
For instance, if $i=1$ then $j=2$ and vice versa.

Now, let us return to Eqs.\ (\ref{Vks1}) and (\ref{Aks1}).
The quantities $V_{{\pmb k}\sigma}^{(i)}$ 
and $A_{{\pmb k}\sigma}^{(i)}$, 
entering the kinetic equations (\ref{kineq1})--(\ref{kineq4}),
depend on the effective potentials 
$V_{\rm eff}^{(i)}$ and ${\pmb A}_{\rm eff}^{(i)}$
in exactly the same way as 
in the absence of Landau quasiparticle interaction 
[when $f^{ij}({\pmb k}, {\pmb k}')=0$].
The only difference is that 
in all equations 
one should replace the bare mass $m_i$ with
the effective mass $m_i^\ast$, or, equivalently,
replace $m_i$ with $p_{{\rm F}i}/v_{{\rm F}i}$.
Moreover,
as follows from Eqs.\ (\ref{Veff1}) and (\ref{Aeff1}),
the gauge transformation properties of 
$V_{\rm eff}^{(i)}$ and ${\pmb A}_{\rm eff}^{(i)}$
coincide with that 
of, respectively, $V$ and ${\pmb A}$ 
[see Eqs.\ (\ref{gaugeV1}) and (\ref{gaugeA1})].
Consequently, the relation between $\delta n_i$, ${\pmb j}_i$ 
and $V_{\rm eff}^{(i)}$, ${\pmb A}_{\rm eff}^{(i)}$ 
in an {\it arbitrary} gauge
is given by the same expressions as for 
a noninteracting Fermi-liquid 
[compare with Eqs.\ (\ref{dn_arbitrary2}) 
and (\ref{j_arbitrary2})], 
\begin{eqnarray}
\delta n_i &=& \Pi_{00}^{(i)} \, 
\left( V_{\rm eff}^{(i)} -\frac{\omega}{q} 
\, A_{{\rm eff}, {\rm l}}^{(i)}\right),
\label{dni2}\\
{\pmb j}_i &=&  \Pi_{\rm tr}^{(i)} \,\,
{\pmb A}_{\rm eff, tr}^{(i)}
+\frac{\pmb q}{\omega} \,\, \Pi_{\rm l}^{(i)} 
\left( V_{\rm eff}^{(i)}  
-\frac{\omega}{q} A_{{\rm eff}, {\rm l}}^{(i)}\right).
\label{ji2}
\end{eqnarray}
%
%
Let us emphasize once again that here the quantities 
$\Pi_{00}^{(i)}$, $\Pi_{\rm l}^{(i)}$, and $\Pi_{\rm tr}^{(i)}$
should be understood as the functions 
of $p_{{\rm F}i}$ and $m_i^\ast$ 
(or $p_{{\rm F}i}$ and $v_{{\rm F}i}=p_{{\rm F}i}/m_i^\ast$)
rather than the functions of $p_{{\rm F}i}$ and $m_i$.
These two sets of variables are equivalent only 
for noninteracting Fermi-liquid, 
when $m_i^{\ast}=m_i$. 

Now we are 
ready
to find the polarization functions
$P_{00}^{(i)}$, $P_{\rm l}^{(i)}$, and $P_{\rm tr}^{(i)}$
for a two-component, strongly interacting Fermi-liquid.
For that, we compare the general expressions 
(\ref{dn_arbitrary}) and (\ref{j_arbitrary}) 
for $\delta n_i$ and ${\pmb j}_i$ 
with the corresponding equations (\ref{dni2}) and (\ref{ji2}).
Taking into account, 
that $V_{\rm eff}^{(i)}$ and ${\pmb A}_{\rm eff}^{(i)}$ 
are given by Eqs.\ (\ref{Veff1}) and (\ref{Aeff1}), 
we obtain
\begin{eqnarray}
P_{00}^{(i)} &=& \frac{\Pi_{00}^{(i)} \left( 
e_i e_j + e_{j} \, \chi_{i j} \, \Pi_{00}^{(j)} 
- e_{i} \, \chi_{j j} \, \Pi_{00}^{(j)} 
\right)}{e_i e_j - e_{i} \, \chi_{j j} \, \Pi_{00}^{(j)}
- e_j \, \chi_{i i} \, \Pi_{00}^{(i)} 
- \chi_{i j} \, \chi_{j i} 
\, \Pi_{00}^{(i)} \, \Pi_{00}^{(j)} 
+ \chi_{i i} \, \chi_{j j} 
\, \Pi_{00}^{(i)} \, \Pi_{00}^{(j)}},
\label{P00}\\
P_{\rm tr}^{(i)} &=& \frac{\Pi_{\rm tr}^{(i)} \left( 
e_i e_j + e_{j} \, \gamma_{i j} \, \Pi_{\rm tr}^{(j)} 
- e_{i} \, \gamma_{j j} \, \Pi_{\rm tr}^{(j)} 
\right)}{e_i e_j - e_{i} \, \gamma_{j j} \, \Pi_{\rm tr}^{(j)}
- e_j \, \gamma_{i i} \, \Pi_{\rm tr}^{(i)} 
- \gamma_{i j} \, \gamma_{j i} 
\, \Pi_{\rm tr}^{(i)} \, \Pi_{\rm tr}^{(j)} 
+ \gamma_{i i} \, \gamma_{j j} 
\, \Pi_{\rm tr}^{(i)} \, \Pi_{\rm tr}^{(j)}},
\label{Ptr}
\end{eqnarray}
while $P_{\rm l}^{(i)}$ can be found from Eq.\ (\ref{PlP00}).
Here 
$\chi_{ij}=f_0^{ij}- (\omega^2/q^2) \, \gamma_{ij}$
and the indices $i$ and $j$ 
refer to different particle species, $i \neq j$.
Eqs.\ (\ref{P00}) and (\ref{Ptr}) 
will be analyzed in a subsequent section.

Here it is convenient to make 
a few comments, concerning the scheme of calculation
of the polarization functions, suggested above.
First, in principle it would be possible to extend this scheme 
taking into account the Landau parameters $f_l^{ij}$ with $l \geq 2$.
However, for that one needs to know 
how the `noninteracting' system 
[with $f^{ij}({\pmb k}, {\pmb k}')=0$] 
responds to a general perturbation 
(not just to the electromagnetic field). 
This is a complex problem without any 
simple solution, such as in the case of $l<2$.

Second, the approach described above can be easily 
generalized to calculate the axial-vector response of the system.
For that one needs to introduce the spin-dependent part 
of the Landau quasiparticle interaction in the kinetic equation (\ref{kineq}).
For a one-component superfluid Fermi-liquid, 
the axial-vector polarization functions were
calculated by Leinson \cite{leinson09}
under the simplified assumption that only nonzero
Landau parameters are $g_0$ and $g_1$
(see Ref.\ \cite{leinson09} for the definition of $g_0$ and $g_1$).
The generalization of his results to the case of mixtures 
is straightforward.
The more difficult problem would be to estimate the effect of 
{\it tensor} quasinucleon interactions 
on the axial-vector response of the system.
We hope to address this problem in our subsequent publication.

\subsection{Various limiting cases 
for $P_{00}^{(i)}$ and $P_{\rm tr}^{(i)}$} 

We consider first a {\it one-component} Fermi-liquid.
The particle species indices can then be suppressed.
The polarization function $P_{00}$ 
for a one-component Fermi-liquid 
was studied by Leggett \cite{leggett66}
in the limit of $q v_{\rm F} \ll \Delta$ and $\omega \ll \Delta$.
Both polarization functions $P_{00}$ and $P_{\rm tr}$ 
were studied in the recent paper by Leinson \cite{leinson09} 
at arbitrary $q v_{\rm F} \ll \mu$ and $\omega \ll \mu$.
In his analysis, Leinson took into account 
only the harmonic $V_0$ of the 
pairing interaction and assumed that $V_{l}=0$ for $l>0$. 
Our results will be compared with 
the results of these two authors.

For a one-component Fermi-liquid
Eqs.\ (\ref{P00}) and (\ref{Ptr}) 
are essentially simplified
\begin{eqnarray}
P_{00} &=& \frac{e \, \Pi_{00}}{e - \chi \, \Pi_{00}},
\label{P00_1}\\
P_{\rm tr} &=& \frac{e \, \Pi_{\rm tr}}{e- \gamma \, \Pi_{\rm tr}},
\label{Ptr_1}
\end{eqnarray}
with
\begin{eqnarray}
\gamma &=& -\frac{m m^\ast}{k_{\rm F}^2} \, f_1, 
\label{gamma_1}\\
\chi   &=& f_0 - \frac{\omega^2}{q^2} \,\gamma.
\label{chi_1}
\end{eqnarray}
To obtain Eqs.\ (\ref{P00_1}) and (\ref{Ptr_1}) 
we put $f_0^{ij}=0$ and $f_1^{ij}=0$ for $i \neq j$ 
in Eqs.\ (\ref{P00}) and (\ref{Ptr}) 
and then suppress the particle species indices.

Eq.\ (\ref{P00_1}) coincides 
with Eq.\ (68) of Leggett \cite{leggett66} 
and with Eq.\ (63) of Leinson \cite{leinson09}
in the limit of $q v_{\rm F} \ll \Delta$ 
and $\omega \ll \Delta$.
In the other limiting case, 
when $q v_{\rm F} \ll \Delta$ and $\omega > 2 \Delta$,
we reproduce the result of Leinson \cite{leinson09}, 
see his Eq.\ (82).
However, at arbitrary $q v_{\rm F}$ and $\omega$ 
our Eq.\ (\ref{P00_1}) differs 
from the general equation (55) for $P_{00}$,
suggested by Leinson \cite{leinson09}.

His Eq.\ (55) 
depends on {\it two} complicated combinations of integrals, $Q$ and $P$, 
that can be easily expressed through each other only 
in the limits discussed above.
In contrast, our Eq.\ (\ref{P00_1}) 
depends solely on the polarization function 
$\Pi_{00}(p_{\rm F}, v_{\rm F})$ 
of noninteracting one-component Fermi-liquid, 
which is the {\it direct consequence} of gauge invariance
of the quantities $\delta n$ and ${\pmb j}$.
%
%
%

Now let us examine Eq.\ (\ref{Ptr_1}) for $P_{\rm tr}$.
Knowledge of $P_{\rm tr}$ allows one
to calculate the transverse-current autocorrelation function $K_{T}$,
\begin{equation}
K_T = \frac{P_{\rm tr}}{e} + \frac{n}{m}.
\label{Kt}
\end{equation}
This quantity was derived by Leinson 
(see Eq.\ (86) of Ref.\ \cite{leinson09}).
Notice that his Eq.\ (86) contains a misprint \cite{leinson10};
one should replace $V_F$ with $p_{\rm F}/m$ in this equation. 
After correcting the misprint, Eq.\ (86) of Leinson coincides 
with our Eq.\ (\ref{Kt}). 
%
%
%

To further check Eq.\ (\ref{Ptr_1}),
we look at the static limit of $P_{\rm tr}$, 
assuming that $\omega=0$ and $q v_{\rm F} \ll \Delta$.
In the static limit 
the particle current density ${\pmb j}$ 
is generated solely by a motion of superfluid liquid component 
(i.e., the normal component is at rest).
It is given by \cite{lp80}
\begin{equation}
{\pmb j} = \frac{\rho_{\rm s}}{m} \,\,  {\pmb V}_{\rm s},
\label{j_phenomenology}
\end{equation}
where $\rho_{\rm s}$ is the superfluid density 
and ${\pmb V}_{\rm s}$ is the superfluid velocity.
The velocity ${\pmb V}_{\rm s}$ depends 
on the gauge-invariant combination of the phase $\varphi$ 
of the Cooper-pair condensate wave function 
and on the electromagnetic potential ${\pmb A}$ 
(see, e.g., Ref.\ \cite{lp80} 
and Sec.\ IV for more details),
\begin{equation}
{\pmb V}_{\rm s} = \frac{1}{2m} \, ({\pmb \nabla} \varphi -2 e {\pmb A}).
\label{Vs}
\end{equation}
In the transverse gauge in which $A_{\rm l}=({\pmb q} {\pmb A})/q=0$,
the phase $\varphi=0$, because it can only depend on 
the scalar $({\pmb q} {\pmb A})=0$. 
Then it follows 
from Eqs.\ (\ref{j_phenomenology}) 
and (\ref{Vs}) that
\begin{equation}
{\pmb j} = - \frac{e \rho_{\rm s}}{m^2} 
\,\, {\pmb A}_{\rm tr}.
\label{static1}
\end{equation}
For a one-component {\it noninteracting} Fermi-liquid 
\begin{equation}
\rho_{\rm s} = m \, n \, \left[1- \Phi(T) \right],
\label{rhos}
\end{equation}
where $\Phi(T)$ is a function of temperature 
[for more details see, e.g., Ref.\ \cite{leggett65}, 
where this function was denoted by $f(T)$].
Using Eqs.\ (\ref{j_arbitrary2}), (\ref{static1}), 
and (\ref{rhos}), one obtains
\begin{equation}
\Pi_{\rm tr}(p_{\rm F}, v_{\rm F}) 
= - \frac{e \,n \, v_{\rm F} \,\left(1- \Phi \right)}{p_{\rm F}}.
\label{Pi_tr}
\end{equation}
This equation together with 
Eqs.\ (\ref{effmass}) and (\ref{Ptr_1}) gives
\begin{equation}
P_{\rm tr} = -\frac{e \, n \, (1- \Phi)}{m \, (1+ F_1 \, \Phi /3)}, 
\label{Ptr_2}
\end{equation}
where $F_1=(m^\ast p_{\rm F}/\pi^2) \, f_1$.
Comparing Eqs.\ (\ref{static1}) and (\ref{Ptr_2}), 
one can determine the superfluid density 
$\rho_{\rm s}$ for {\it interacting} Fermi-liquid,
\begin{equation}
\rho_{\rm s} = - \frac{m^2}{e} \, P_{\rm tr}=
\frac{m\, n \, \left[1- \Phi(T)\right]}{1+ F_1 \, \Phi(T) /3}.
\label{rhos_2}
\end{equation}
It coincides with the corresponding Eq.\ (72) of Leggett \cite{leggett65}.
(Notice that, Leggett calculated the `normal' density 
$\rho_{\rm n} \equiv m n -\rho_{\rm s}$.)
Thus, we demonstrate that in the static limit 
our Eq.\ (\ref{Ptr_1}) 
reproduces the well-known result of Ref.\ \cite{leggett65}.

We now turn our attention to the two-component Fermi-liquid
and discuss first the static limit ($\omega=0$ and $q v_{{\rm F} i}\ll \Delta_j$).
To obtain $P_{\rm tr}^{(i)}$ in this limit one should use
Eq.\ (\ref{Ptr}) with $\Pi_{\rm tr}^{(i)}$ given by expression, 
similar to Eq.\ (\ref{Pi_tr}),
\begin{equation}
\Pi_{\rm tr}^{(i)}(p_{{\rm F}i}, v_{{\rm F}i}) 
= - \frac{e_i \,n_i \, v_{{\rm F}i} \,\left(1- \Phi_i \right)}{p_{{\rm F}i}},
\label{Pi_tr1}
\end{equation}
where $\Phi_i$ is the same function of temperature as $\Phi$ 
(for more details, see Ref.\ \cite{gh05}).

On the other hand, 
in the static limit 
the general hydrodynamics 
of superfluid mixtures is applicable, 
which states that in the absence of normal current
(see, e.g., Ref.\ \cite{ab75}),
\begin{equation}
{\pmb j}_i = \sum_{j} \, \frac{\rho_{ij}}{m_i} \,\,  {\pmb V}_{{\rm s}j}.
\label{j_phenomenology2}
\end{equation}
Here $\rho_{ij}$ is the entrainment matrix 
(also termed the Andreev-Bashkin or mass-density matrix) and 
${\pmb V}_{{\rm s }i}$ is the superfluid 
velocity for $i$-th particle species.
It can be expressed through the phase $\varphi_i$ 
by an equation, similar to Eq.\ (\ref{Vs}),
\begin{equation}
{\pmb V}_{{\rm s}i} = \frac{1}{2m_i} \, ({\pmb \nabla} \varphi_i -2 e_i {\pmb A}).
\label{Vs_2}
\end{equation}
Again, as for a one-component liquid, 
in the transverse gauge $\varphi_i=0$ 
and from Eqs.\ (\ref{j_phenomenology2}) and (\ref{Vs_2})
we have
\begin{equation}
{\pmb j}_i =-\left( 
\frac{e_i \, \rho_{i i}}{m_{i}^2} 
+ \frac{e_j \, \rho_{ij}}{m_{i} \, m_{j}} 
\right) \, {\pmb A}_{\rm tr},
\label{static2}
\end{equation}
or, in view of Eq.\ (\ref{j_arbitrary}),
\begin{equation}
P_{\rm tr}^{(i)} = -\left( 
\frac{e_i \, \rho_{i i}}{m_{i}^2} 
+ \frac{e_j \, \rho_{ij}}{m_{i} \, m_{j}} 
\right).
\label{Ptr_3}
\end{equation}
In Eqs.\ (\ref{static2}) and (\ref{Ptr_3}) indices $i$ and $j$ refer
to different particle species, $i \neq j$.
Comparing Eq.\ (\ref{Ptr_3}) with Eq.\ (\ref{Ptr}) 
and taking into account Eq.\ (\ref{Pi_tr1}), 
one can determine the entrainment matrix $\rho_{ij}$
and verify that it coincides with the result of Ref.\ \cite{gh05},
obtained in a quite different way.
(It should be noted that in Ref.\ \cite{gh05} 
it is additionally demonstrated 
that the higher harmonics $f^{ij}_l$ with $l \geq 2$ 
do not contribute to $\rho_{ij}$.)

The next interesting limiting case 
is realized if one particle species, say, $i=1$, 
is charged while the other is not ($e_2=0$).
It follows then from general 
equation (\ref{P00}) 
that 
\begin{eqnarray}
P^{(1)}_{00} &=& \frac{e_1 \, \widetilde{\Pi}^{(1)}_{00} \,  
\left(   1- \chi_{22} \, \widetilde{\Pi}^{(2)}_{00}   \right)}
{1 - \chi_{22} \, \widetilde{\Pi}^{(2)}_{00} -
\chi_{11} \, \widetilde{\Pi}^{(1)}_{00} - \chi_{12} \, \chi_{21} \,
\widetilde{\Pi}^{(1)}_{00} \, \widetilde{\Pi}^{(2)}_{00}
+\chi_{11} \, \chi_{22} \,
\widetilde{\Pi}^{(1)}_{00} \, \widetilde{\Pi}^{(2)}_{00} },
\label{P001}\\
P^{(2)}_{00} &=& \frac{e_1 \, \chi_{21} 
\, \widetilde{\Pi}^{(1)}_{00} \, \widetilde{\Pi}^{(2)}_{00}}
{1 - \chi_{22} \, \widetilde{\Pi}^{(2)}_{00} -
\chi_{11} \, \widetilde{\Pi}^{(1)}_{00} - \chi_{12} \, \chi_{21} \,
\widetilde{\Pi}^{(1)}_{00} \, \widetilde{\Pi}^{(2)}_{00}
+\chi_{11} \, \chi_{22} \,
\widetilde{\Pi}^{(1)}_{00} \, \widetilde{\Pi}^{(2)}_{00} }.
\label{P002}
\end{eqnarray}
Here the function $\widetilde{\Pi}^{(i)}_{00}$ 
is independent of the electric charge $e_i$, 
$\widetilde{\Pi}^{(i)}_{00} \equiv \Pi^{(i)}_{00}/e_i$.
An analogous expressions
for $P_{\rm tr}^{(1)}$ and $P_{\rm tr}^{(2)}$
can be obtained with the help of Eq.\ (\ref{Ptr}).
One sees from Eqs.\ (\ref{P001}) and (\ref{P002})
[and from the corresponding equations for $P_{\rm tr}^{(i)}$]
that neutral particles not only modify 
the polarization function $P_{00, \, {\rm tr}}^{(1)}$ 
of charged particles, 
but also respond themselves to electromagnetic field, 
because $P_{00, \, {\rm tr}}^{(2)} \neq 0$.
However, the neutral particles do not contribute to 
the dielectric functions of the liquid,
since, as follows from Eqs.\ (\ref{e_l}) and (\ref{e_tr}), 
$\varepsilon_{\rm l}$ and $\varepsilon_{\rm tr}$ are given by 
\begin{eqnarray}
\varepsilon_{\rm l} &=& 
1- \frac{4 \pi}{q^2}  \,\, e_1 \, P_{00}^{(1)},
\label{e_l2}\\
\varepsilon_{\rm tr} &=& 
1 - \frac{4 \pi}{\omega^2} \,\, e_1 \, P_{\rm tr}^{(1)}.
\label{e_tr2}
\end{eqnarray}

Finally, we mention another interesting property 
of the general solution (\ref{P00}) and (\ref{Ptr}).
Assume that the polarization functions 
$\Pi_{00}^{(i=1, \,2)}$ [or $\Pi_{\rm tr}^{(i=1, \,2)}$] 
of noninteracting Fermi-liquid are small so that one can neglect 
the terms of the order of 
$\Pi_{00}^{(i)}\, \Pi_{00}^{(j)}$ 
[or $\Pi_{\rm tr}^{(i)}\, \Pi_{\rm tr}^{(j)}$].
Then, as follows from Eqs.\ (\ref{P00}) and (\ref{Ptr}),
in the linear approximation,
\begin{equation}
P_{00}^{(i)} \approx \Pi_{00}^{(i)} \quad \quad \quad
\left[ {\rm or} \,\, P_{\rm tr}^{(i)} \approx\Pi_{\rm tr}^{(i)} \right].
\label{true}
\end{equation}
That is, the polarization functions 
are {\it not} modified by the first two harmonics 
$f_0^{ij}$ and $f_1^{ij}$ 
of the Landau quasiparticle interaction.
In the next section we demonstrate
that 
in some cases 
this conclusion
is correct even 
if we take into account {\it all} harmonics 
$f_l^{ij}$ with $l \geq 0$.
 
\subsection{Transverse polarization function in the Pippard limit} 

In the Pippard limit we have 
$q v_{{\rm F} j} \gg \Delta_i$, $q v_{{\rm F} j} \gg \omega$,
and $q v_{{\rm F} j} \ll \mu_i$.
It is especially important to know the polarization functions 
$P_{00}^{(i)}$ and $P_{\rm tr}^{(i)}$
in this limit because they are required, 
for instance, for calculating the kinetic coefficients 
of a multi-fluid Fermi-mixture 
in neutron-star cores \cite{sy07,sy08}. 

In the first approximation, 
$P_{00}^{(i)}$ 
does not depend on the gap $\Delta_i$ 
and the frequency $\omega$
and agrees with the corresponding expression for normal matter, 
describing the ordinary static screening of particles.
It
can be easily obtained from Eq.\ (\ref{P00}) 
if we notice that
for a normal one-component noninteracting Fermi-liquid
one has (see, e.g., Ref.\ \cite{pn66})
\begin{equation}
\Pi_{00}^{(i)} = - \frac{e_i \, p_{{\rm F}i}^2}{\pi^2 \, v_{{\rm F}i}}.
\label{Pi00a}
\end{equation}
Strictly speaking, Eq.\ (\ref{P00}) that we employ,
was derived under the assumption
that the only harmonics $f_0^{ij}$ and $f_1^{ij}$ 
of the Landau quasiparticle interaction are nonzero. 
However, one can easily verify
that, for a {\it normal} Fermi-liquid, 
$P_{00}^{(i)}$ is still given (at small ${\pmb q}$ and $\omega$) 
by Eq.\ (\ref{P00})
even if we allow for higher harmonics 
$f_{l}^{ij}$ with $l \geq 2$.
For a one-component Fermi-liquid 
this was demonstrated, for example, in Ref.\ \cite{pn66}. 

Now let us consider the transverse 
polarization function $P_{\rm tr}^{(i)}$.
In the Pippard limit the polarization function 
$\Pi_{\rm tr}^{(i)}$ is small, 
$\Pi_{\rm tr}^{(i)}
= O[\Delta_i/(q v_{{\rm F}i}) + \omega/(q v_{{\rm F}i})]$.
It follows then, from the discussion 
at the end of the preceding section,
that the first two harmonics $f_0^{ij}$ and $f_1^{ij}$ 
of the Landau quasiparticle interaction
have no influence on $P_{\rm tr}^{(i)}$, 
so that it is given by Eq.\ (\ref{true}).

Below we demonstrate that this result remains correct even 
if we take into account other harmonics 
$f_l^{ij}$ with $l \geq 2$.
In the static limit ($\omega=0$)
this was first shown by Leggett \cite{leggett65}.
More precisely, we prove that 
the transverse polarization function for
a noninteracting system 
coincides with the function
for a system with an arbitrary harmonic $f_l^{ij}$ 
($l \geq 2$) switched on.

We consider first the simplified situation in which 
the pairing potential 
$V^{(i)}({\pmb k}, {\pmb k}')$ is a constant; 
the generalization of our results 
to the case of an arbitrary 
$V^{(i)}({\pmb k}, {\pmb k}')$
is briefly discussed at the end of the present section. 
In other words, we take into account 
only the first term $V_{0}^{(i)}$ 
in the expansion (\ref{Vkk}) 
of $V^{(i)}({\pmb k}, {\pmb k}')$ 
in Legendre polynomials
and neglect all other terms, 
$V_l^{(i)}=0$ for $l \geq 1$.
In that case the functions 
$O_{{\pmb k}\sigma +}^{(i)}$ and $O_{{\pmb k}\sigma -}^{(i)}$
in Eqs.\ (\ref{kineq1})--(\ref{kineq4})
are some constants 
depending on the scalars 
${\pmb q} {\pmb A}$ and $V$
(see, e.g., Ref.\ \cite{lp80}).
In the transverse gauge, 
where ${\pmb q} {\pmb A} = {\pmb q} {\pmb A}_{\rm tr} = 0$, 
and in the absence 
of scalar electromagnetic potential $V$, 
they vanish, 
$O_{{\pmb k}\sigma +}^{(i)}=O_{{\pmb k}\sigma -}^{(i)}=0$.
The solution to the system 
of equations (\ref{kineq1})--(\ref{kineq4})
is then simplified and we have for 
$\delta n_{{\pmb k} \sigma +}^{(i)}({\pmb q}, \omega, \Delta_i)$ 
and $\delta n_{{\pmb k} \sigma -}^{(i)}({\pmb q}, \omega, \Delta_i)$
\begin{eqnarray}
\delta n_{{\pmb k} \sigma +}^{(i)}({\pmb q}, \omega, \Delta_i)
&=&  \left[ {\pmb q}{\pmb v}_i \, A_{{\pmb k}\sigma}^{(i)} 
-\omega \, V_{{\pmb k}\sigma}^{(i)} \right] \frac{M_1}{D},
\label{dnks+2}\\
\delta n_{{\pmb k} \sigma -}^{(i)}({\pmb q}, \omega, \Delta_i) 
&=& \omega \, A_{{\pmb k}\sigma}^{(i)} \,\, \frac{ M_1 }{D}
+ V_{{\pmb k}\sigma}^{(i)} \,\, \frac{M_2}{D},
\label{dnks-2}
\end{eqnarray}
where 
\begin{eqnarray}
M_1 &=& \frac{1}{2} \, 
{\pmb q}{\pmb v}_i \, \left[ 4 \xi_{\pmb k}^{(i)2} - \omega^2 \right]\,
\left[ F_{{\pmb k}_+}^{(i)} + F_{{\pmb k}_-}^{(i)} \right]
\nonumber\\
&+& \xi_{\pmb k}^{(i)} \left[ 4 \xi_{\pmb k}^{(i)2} + 4 \Delta_i^2 -\omega^2 \right]\,
\left[ F_{{\pmb k}_+}^{(i)} - F_{{\pmb k}_-}^{(i)} \right],
\label{M1}\\
M_2 &=& \frac{1}{2} \, 
\left[ 4 \Delta_i^2 \, \omega^2 - 4 ({\pmb q} {\pmb v}_i)^2 \, \xi_{\pmb k}^{(i)2}
+ ({\pmb q} {\pmb v}_i)^2 \, \omega^2 \right]\,
\left[ F_{{\pmb k}_+}^{(i)} + F_{{\pmb k}_-}^{(i)} \right]
\nonumber\\
&+& {\pmb q}{\pmb v}_i \, \xi_{\pmb k}^{(i)} \, 
\left[4 \xi_{\pmb k}^{(i)2} - \omega^2 \right] \,
\left[ F_{{\pmb k}_-}^{(i)} - F_{{\pmb k}_+}^{(i)} \right],
\label{M2}\\
D &=& \left(E_{{\pmb k}_-}^{(i)}- E_{{\pmb k}_+}^{(i)} -\omega - i0 \right)
 \left(E_{{\pmb k}_-}^{(i)} + E_{{\pmb k}_+}^{(i)} -\omega -i0 \right)
 \nonumber\\
 &\times&
\left(E_{{\pmb k}_-}^{(i)} - E_{{\pmb k}_+}^{(i)} +\omega + i0 \right)
\left(E_{{\pmb k}_-}^{(i)} + E_{{\pmb k}_+}^{(i)} +\omega +i0 \right).
\label{D}
\end{eqnarray}
In Eqs.\ (\ref{dnks+2}) and (\ref{dnks-2}) 
$V_{{\pmb k}\sigma}^{(i)}$ and $A_{{\pmb k}\sigma}^{(i)}$
are the smooth functions of ${\pmb k}$, 
defined in Eqs.\ (\ref{Vks}) and (\ref{Aks}), respectively.
For our problem they can be rewritten as
\begin{eqnarray}
V_{{\pmb k} \sigma}^{(i)} &=& \sum_{{\pmb k}' \sigma' j} 
f^{ij}_l \, P_l(\cos \theta) \,\, 
\delta n_{{\pmb k}' \sigma' -}^{(j)}({\pmb q}, \omega, \Delta_j),
\label{Vks2}\\
A_{{\pmb k} \sigma}^{(i)} &=& 
2 \alpha_i \,\, \frac{{\pmb k} {\pmb A}}{m_i} 
-\sum_{{\pmb k}' \sigma' j} 
f^{ij}_l \, P_l(\cos \theta) \,\, 
\delta n_{{\pmb k}' \sigma' +}^{(j)}({\pmb q}, \omega, \Delta_j),
\label{Aks2}
\end{eqnarray}
where $\theta$ is the angle between ${\pmb k}$ and ${\pmb k}'$ and $l \geq 2$.

As already mentioned above, 
Leggett \cite{leggett65} showed 
that the static function $P_{\rm tr}^{(i)}({\pmb q},0, \Delta_i)$
is not affected by 
the Landau quasiparticle interaction.
Therefore, it is sufficient to analyze the difference
$P_{\rm tr}^{(i)}({\pmb q}, \omega, \Delta_i) 
- P_{\rm tr}^{(i)}({\pmb q}, 0, \Delta_i)$
and prove that it is independent of $f_l^{ij}$.
Using Eq.\ (\ref{Ji2}), we obtain
\begin{eqnarray}
{\pmb j}_i({\pmb q}, \omega, \Delta_i) - {\pmb j}_i({\pmb q}, 0, \Delta_i)
&=& \left[ P_{\rm tr}^{(i)}({\pmb q}, \omega, \Delta_i) 
- P_{\rm tr}^{(i)}({\pmb q}, 0, \Delta_i) \right] {\pmb A}_{\rm tr} 
\nonumber\\
&=& \frac{1}{2} \, \sum_{{\pmb k} \sigma} \frac{\pmb k}{m_i} 
\left[ \delta n_{{\pmb k} \sigma +}^{(i)}({\pmb q}, \omega, \Delta_i) 
- \delta n_{{\pmb k} \sigma +}^{(i)}({\pmb q}, 0, \Delta_i) \right].
\label{Ji3}
\end{eqnarray}
The function 
$\delta n_{{\pmb k} \sigma +}^{(i)}({\pmb q}, \omega, \Delta_i) 
- \delta n_{{\pmb k} \sigma +}^{(i)}({\pmb q}, 0, \Delta_i)$
is nonzero only in a narrow region near the Fermi surface, 
when $k \sim k_{{\rm F}i}$.
Furthermore, because of the denominator $D$ [see Eq.\ (\ref{D})], 
this function has a sharp maximum at 
${\pmb q}{\pmb v}_i \la (\omega+\Delta_j)$.
Introducing the longitudinal 
${\pmb k}_{\rm l} \, \| \, {\pmb q}$ and 
transverse  ${\pmb k}_{\rm tr} \, \bot \, {\pmb q}$
components of the vector 
${\pmb k}={\pmb k}_{\rm l}+{\pmb k}_{\rm tr}$,
this inequality can be rewritten as
\begin{equation}
k_{\rm l} \la \frac{m_i (\omega + \Delta_i)}{q} 
\ll k_{\rm tr} \sim k_{{\rm F}i}.
\label{ineq}
\end{equation}
The main contribution to the integral (\ref{Ji3})
comes from $k_{\rm l}$ satisfying Eq.\ (\ref{ineq}).
Keeping this in mind, it is straightforward to verify
(see, e.g., Ref.\ \cite{lp81}) 
that for noninteracting Fermi-liquid one has 
\begin{equation}
{\pmb j}_i({\pmb q}, \omega, \Delta_i) 
- {\pmb j}_i({\pmb q}, 0, \Delta_i) = 
O\left( \frac{\omega+\Delta_i}{q v_{{\rm F}i}}\right).
\label{noninteract}
\end{equation}

Now let us analyze whether 
the Landau quasiparticle interaction influences this result.
For this purpose we
inspect the terms in the function 
$\delta n_{{\pmb k} \sigma +}^{(i)}({\pmb q}, \omega, \Delta_i) 
- \delta n_{{\pmb k} \sigma +}^{(i)}({\pmb q}, 0, \Delta_i)$
which depend on $f^{ij}_l$.
From Eqs.\ (\ref{dnks+2}), (\ref{Vks2}), and (\ref{Aks2}) 
it follows that they have the form,
\begin{equation}
I=\sum_{{\pmb k}' \sigma'} 
f^{ij}_l \, 
P_l\left(\frac{{\pmb k} {\pmb k}'}{k_{{\rm F}i} k_{{\rm F}j}} \right) 
\,\, \delta n_{{\pmb k}' \sigma' +}^{(j)}
\quad {\rm and} \quad
II=\sum_{{\pmb k}' \sigma'} 
f^{ij}_l \, 
P_l\left(\frac{{\pmb k} {\pmb k}'}{k_{{\rm F}i} k_{{\rm F}j}} \right) 
\,\, \delta n_{{\pmb k}' \sigma' -}^{(j)}.
\label{terms}
\end{equation}
By demonstrating that these integrals are quadratic in 
$(\omega+\Delta_j)/(q v_{{\rm F}i})$ 
we prove that $P_{\rm tr}^{(i)}
=\Pi_{\rm tr}^{(i)}(p_{{\rm F}i}, v_{{\rm F}i})$ 
in the Pippard limit.
Below we consider in detail 
the first term in Eq.\ (\ref{terms});
the analysis of the second term is similar.

The first term can be presented as
\begin{eqnarray}
I &=& \sum_{{\pmb k}' \sigma'} 
f^{ij}_l \, 
P_l\left(\frac{{\pmb k} {\pmb k}'}{k_{{\rm F}i} k_{{\rm F}j}} \right) 
\,\, \left[\delta n_{{\pmb k}' \sigma' +}^{(j)}({\pmb q}, \omega, \Delta_j) 
- \delta n_{{\pmb k}' \sigma' +}^{(j)}({\pmb q}, 0, \Delta_j) \right] 
\nonumber\\
&+& \sum_{{\pmb k}' \sigma'} 
f^{ij}_l \, 
P_l\left(\frac{{\pmb k} {\pmb k}'}{k_{{\rm F}i} k_{{\rm F}j}} \right) 
\,\, \left[\delta n_{{\pmb k}' \sigma' +}^{(j)}({\pmb q}, 0, \Delta_j) 
- \delta n_{{\pmb k}' \sigma' +}^{(j)}({\pmb q}, 0, 0) \right] 
\nonumber\\
&+&\sum_{{\pmb k}' \sigma'} 
f^{ij}_l \, 
P_l\left(\frac{{\pmb k} {\pmb k}'}{k_{{\rm F}i} k_{{\rm F}j}} \right) 
\,\, \delta n_{{\pmb k}' \sigma' +}^{(j)}({\pmb q}, 0, 0).
\label{I}
\end{eqnarray}
One can easily verify 
that the last integral 
here
vanishes at $l \neq 1$ 
and thus can be omitted.
Furthermore, 
because of the same reasons as those
discussed below Eq.\ (\ref{Ji3}),
the main contribution to the first 
two integrals in Eq.\ (\ref{I})
comes from such a region of ${\pmb k}'$, that 
\begin{equation}
k'_{\rm l} \la \frac{m_j (\omega + \Delta_j)}{q} 
\ll k'_{\rm tr} \sim k_{{\rm F}j}.
\label{ineq2}
\end{equation}
From symmetry arguments it follows that 
the functions in square brackets in these integrals
can generally be written in the form
\begin{equation}
[\ldots]
= ({\pmb k}' {\pmb A}) \,\, G\left({\pmb k}' {\pmb q}, \,\, k' \right),
\label{symm_arg}
\end{equation}
where $G({\pmb k}' {\pmb q}, \,\,k'^2)$ is a scalar function.

Now, using Eqs.\ (\ref{ineq}), (\ref{ineq2}), and (\ref{symm_arg}),
one may write
\begin{eqnarray}
&&\sum_{{\pmb k}' \sigma'} 
f^{ij}_l \, 
P_l\left(\frac{{\pmb k} {\pmb k}'}{k_{{\rm F}i} k_{{\rm F}j}} \right) 
\,\, \left[\ldots \right] =  \sum_{{\pmb k}' \sigma'} 
f^{ij}_l \, 
P_l\left(\frac{k_{\rm l}k'_{\rm l} + 
{\pmb k}_{\rm tr} {\pmb k}'_{\rm tr}}{k_{{\rm F}i} k_{{\rm F}j}} \right)
({\pmb k}' {\pmb A}) \,\, G\left({\pmb k}' {\pmb q}, \,\, k' \right)
\nonumber\\
&=& \sum_{ k'_{\rm l}\, {\pmb k}'_{\rm tr} \, \sigma'} 
f^{ij}_l \, 
P_l\left(\frac{{\pmb k}_{\rm tr} {\pmb k}'_{\rm tr}}
{k_{{\rm F}i} k_{{\rm F}j}} \right)
({\pmb k}'_{\rm tr} {\pmb A}_{\rm tr}) \,\, 
G\left(k'_{\rm l} q, \,\, k' \right) 
+ O\left( \frac{(\omega + \Delta_i)}{q v_{{\rm F}i}} 
\frac{(\omega + \Delta_j)}{q v_{{\rm F}j}} \right)
\nonumber\\ 
&=& O\left( \frac{(\omega + \Delta_i)}{q v_{{\rm F}i}} 
\frac{(\omega + \Delta_j)}{q v_{{\rm F}j}} \right).
\label{Iint}
\end{eqnarray}
Here we used the fact that the integral
over directions of ${\pmb k}'_{\rm tr}$
vanishes for $l \neq 1$.
Thus, we demonstrate that 
$I \sim (\omega+\Delta_i)/( q v_{{\rm F}i})\, \times
(\omega+\Delta_j)/(q v_{{\rm F}j})$ 
and, consequently, 
the Landau quasiparticle interaction
has no influence on $P_{\rm tr}^{(i)}$ in the Pippard limit.

The consideration of this section is simplified since we 
take into account only the harmonic $V_0^{(i)}$ 
of the pairing potential.
However, it seems plausible
(though we have not checked it in detail) 
that the inclusion of other harmonics $V_l^{(i)}$ 
with $l \geq 1$ will not change the result.
In principle, a prove of this more general statement
should be 
similar to the prove presented here, 
but the equations to be analyzed, 
are much more complicated.
In particular, 
the function $\delta n_{{\pmb k} \sigma +}^{(i)}$ 
will depend, in addition, on the integrals 
$O_{{\pmb k} \sigma+}^{(i)}$ 
and $O_{{\pmb k} \sigma -}^{(i)}$ 
[see Eqs.\ (\ref{O+}) and (\ref{O-})].

Finally, let us make 
a comment concerning 
the coefficients $f_1^{ij}$ and $V_1^{(i)}$ 
of the first harmonic.
From the analysis presented above 
it is clear that the coefficient $f_1^{ij}$ 
plays a very special role, 
because the integrals in Eqs.\ (\ref{I}) and (\ref{Iint})
do not vanish at $l=1$.
It might think that the situation 
with the coefficient $V_{1}^{(i)}$ 
of the pairing potential 
is analogous, 
so that the integrals 
\begin{equation}
\sum_{{\pmb k}' \sigma} 
V_1^{(i)} \, P_1(\cos \theta) \, \delta s_{{\pmb k}' \sigma -}^{(i)}
({\pmb q}, \omega, \Delta_i)
\quad {\rm and} \quad
\sum_{{\pmb k}' \sigma} 
V_1^{(i)} \, P_1(\cos \theta) \, \delta s_{{\pmb k}' \sigma +}^{(i)}
({\pmb q}, \omega, \Delta_i)
\label{zero}
\end{equation}
are nonzero. 
However, this is not the case;
they {\it vanish} due to the symmetry relations 
(\ref{dsk+sym}) and (\ref{dsk-sym}).

\section{A phenomenological approach to kinetic equation 
at small ${\pmb q}$ and $\omega$}

In this section we analyze the kinetic equation
in the `quasiclassical' limit,
$q v_{{\rm F}i} \ll \Delta_j$ 
and $\omega \ll \Delta_j$ (see, e.g., Refs.\ \cite{leggett65,bn69}).
This limit is especially important 
for various applications, for instance, 
to study low-frequency long-wavelength collective modes
propagating in superfluid matter and
to calculate kinetic coefficients.

For a one-component Fermi-liquid
the quasiclassical limit of the matrix kinetic equation (\ref{kineq}) 
was thoroughly examined
in Ref.\ \cite{bn69}.
It was demonstrated, 
that Eq.\ (\ref{kineq})
can be substantially simplified
by introducing a concept of Bogoliubov excitations.
In particular, the kinetic equation for Bogoliubov excitations
acquires a {\it scalar}
(rather than matrix) form.
For a two-component Fermi-mixture the analysis 
is quite similar.
Here we do not attempt to perform it;
an interested reader is referred 
to Ref.\ \cite{bn69} for 
more details.
Instead, we follow a more intuitive phenomenological approach
allowing to formulate the kinetic equation 
in the {\it non-linear} regime
(in contrast to 
Ref.\ \cite{bn69}, 
where the kinetic equation 
is studied only in the linear approximation).
We call this regime {\it non-linear} 
in a sense that, for instance, 
it allows us to study the nonequilibrium variations
of the energy gap 
which are comparable to $\Delta_i$.
However, because we use the Landau theory of Fermi-liquids,
we still assume that the quasiparticle distribution
only slightly differs (in the vicinity of the Fermi surface) 
from the step function. 
Our results will be discussed and compared with those
available in the literature in the end of the present section.

In the previous sections 
the quasiparticle momentum was denoted 
as ${\pmb k}$.
In the Appendix A we demonstrate
that in the presence of electromagnetic field 
${\pmb k}$ is actually a {\it generalized} momentum.
To distinguish between ${\pmb k}$ and the real momentum, 
the latter will be denoted
by ${\pmb p}$.
It is more convenient to use {$\pmb p$} 
instead of ${\pmb k}$ 
in the consideration below.

\subsection{Local analysis} 
 
In the quasiclassical limit one can assume that
a Landau quasiparticle (or a Bogoliubov excitation) 
with a certain momentum ${\pmb p}$ possesses, 
at the same time,
a certain coordinate ${\pmb r}$.
Consequently, 
such quantities as
the distribution of Landau quasiparticles 
(Bogoliubov excitations), 
their energy, or the energy gap can be considered 
as functions of
${\pmb p}$ and ${\pmb r}$.
To find how these quantities are related to each other  
it is sufficient 
to analyze the system {\it locally}.

Let us consider a two-component Fermi-liquid 
out of thermodynamic equilibrium.
To simplify the problem we {\it neglect} 
for a while the electromagnetic field, 
assuming that the liquid is composed of neutral particles. 
The electromagnetic effects will be 
taken into account
in Sec.\ IVC.
Our aim will be to calculate the energy density $E$ 
of superfluid matter
in the neighborhood of a point ${\pmb r}$.
In the vicinity of this point the matter
is 
almost homogeneous. 
Thus, it can be approximately described by a
uniform Hamiltonian $H$ 
(see, e.g., Refs.\ \cite{leggett65,gh05,gkh09b}),
\begin{equation}
H - \sum_i \breve{\mu}_i N_i = H_{\rm LF} + H_{\rm pairing},
\label{H}
\end{equation}
where $N_i$ is the number density operator; 
$\breve{\mu}_i$ is the nonequilibrium analogue 
of the chemical potential $\mu_i$
to be determined below;
$H_{\rm LF}$ is the Fermi-liquid Hamiltonian, 
\begin{eqnarray}
{H}_{\rm LF} &=& \sum_{{\pmb p} \sigma i}
\left( \varepsilon_{{\pmb p} \, 0}^{(i)} -\breve{\mu}_{i} \right)
\left( a_{{\pmb p}\sigma}^{(i) \dagger} a_{{\pmb p}\sigma}^{(i)} 
- \theta_{\pmb p}^{(i)} \right)
\nonumber \\
&+& \frac{1}{2}  \sum_{{\pmb p} {\pmb p} '  \sigma \sigma' i j}
f^{i j}({\pmb p}, {\pmb p}') 
\left(a_{{\pmb p}\sigma}^{(i) \dagger} a_{{\pmb p}\sigma}^{(i)} 
- \theta_{\pmb p}^{(i)} \right)
\left(a_{{\pmb p}' \sigma'}^{(j) \dagger} a_{{\pmb p}' \sigma'}^{(j)} 
- \theta_{{\pmb p}'}^{(j)} \right),
\label{LF}
\end{eqnarray}
and $H_{\rm pairing}$ is the pairing Hamiltonian.
In the presence of superfluid currents it is given by 
(see, e.g., Ref.\ \cite{gh05})
\begin{equation}
H_{\rm pairing} = \sum_{{\pmb p} {\pmb p}' i}
V_{{\pmb Q}_i}^{(i)} \left({\pmb p}, {\pmb p'} \right) \,
a_{{\pmb p}' + {\pmb Q}_{i} \uparrow}^{(i) \dagger}
a_{-{\pmb p}' + {\pmb Q}_{i} \downarrow}^{(i) \dagger} \,
a_{-{\pmb p} + {\pmb Q}_{i} \downarrow}^{(i) }
a_{{\pmb p} + {\pmb Q}_{i} \uparrow}^{(i) }.
\label{Hpairing}
\end{equation}
Here, $2 {\pmb Q}_i = 2 m_i V_{{\rm s}i}$ 
is the momentum of a Cooper-pair in the condensate.
It is related by Eqs.\ (\ref{QQ2}) and (\ref{phase2}) 
to the quantity $O_{{\pmb k}\sigma-}^{(i)}$, 
introduced in Sec.\ IIB.
The matrix element $V_{{\pmb Q}_i}^{(i)}({\pmb p}, {\pmb p}')$ 
in Eq.\ (\ref{Hpairing}) 
describes scattering of a pair of Landau quasiparticles from the states 
$({\pmb p}+{\pmb Q}_{i}, \uparrow)$, $(-{\pmb p}+{\pmb Q}_{i},
\downarrow)$ to states $({\pmb p'}+{\pmb Q}_{i}, \uparrow)$,
$(-{\pmb p'}+{\pmb Q}_{i}, \downarrow)$.
In Ref.\ \cite{gh05} it is argued that
$V_{{\pmb Q}_i}^{(i)}({\pmb p}, {\pmb p}') 
\approx V^{(i)}({\pmb p}, {\pmb p}')$ 
[up to small terms $\sim (Q_i/k_{{\rm F}j})^2$]. 

To find the energy density $E$ one needs 
to diagonalize the Hamiltonian (\ref{H}).
To do that, we rewrite Eq.\ (\ref{H}) in terms 
of Bogoliubov operators 
$b_{{\pmb p}\sigma}^{(i)}$, 
defined as (see, e.g., Ref.\ \cite{gh05})
\begin{eqnarray}
a_{{\pmb p} + {\pmb Q}_{i} \uparrow}^{(i)}
&=& u_{\pmb p}^{(i)} \,  b_{{\pmb p} + {\pmb Q}_{i} \uparrow}^{(i)}
+ v_{\pmb p}^{(i)} \, b_{-{\pmb p} + {\pmb Q}_{i} \downarrow}^{(i) \dagger},
\label{bpQ1} \\
a_{{\pmb p} + {\pmb Q}_{i} \downarrow}^{(i)}
&=& u_{\pmb p}^{(i)} \, b_{{\pmb p} + {\pmb Q}_{i} \downarrow}^{(i)}
- v_{\pmb p}^{(i)} \, b_{-{\pmb p} + {\pmb Q}_{i} \uparrow}^{(i) \dagger},
\label{bpQ2}
\end{eqnarray}
where $u_{\pmb p}^{(i)}$ and $v_{\pmb p}^{(i)}$ 
are even functions of ${\pmb p}$, 
\begin{equation}
u_{\pmb p}^{(i)} = u_{- {\pmb p}}^{(i)}, \quad \quad
v_{\pmb p}^{(i)} = v_{- {\pmb p}}^{(i)},
\label{even}
\end{equation}
normalized by the condition 
\begin{equation}
u_{\pmb p}^{(i) 2} + v_{\pmb p}^{(i) 2} =1.
\label{norm}
\end{equation}
Generally, the coefficients $u_{\pmb p}^{(i)}$ and $v_{\pmb p}^{(i)}$ 
are complex. 
However, 
at some moment of time 
they can be chosen to be real 
in the vicinity of our point ${\pmb r}$
by a suitable phase transformation.

Being expressed through the Bogoliubov operators, 
the Hamiltonian (\ref{H}) takes the diagonal form. 
Thus, one gets the following expression for the energy density
\begin{gather}
E- \sum_{i} \breve{\mu}_{i}  n_{i}=
\sum_{{\pmb p} \sigma i}
\left[ 
\varepsilon_{{\pmb p}+{\pmb Q}_{i} \, 0}^{(i)}
-\breve{\mu}_{i} \right]\,
\left( \mN_{{\pmb p}+ {\pmb Q}_{i}}^{(i)} - \theta_{{\pmb p}+{\pmb Q}_i}^{(i)} \right)
\nonumber \\
+ \frac{1}{2}  \sum_{{\pmb p} {\pmb p} '  \sigma \sigma' i j}
f^{i j} \left(\vp + \vQa, \vps + {\pmb Q}_{j} \right)
\left( \mN_{{\pmb p}+ {\pmb Q}_{i}}^{(i)} 
- \theta_{{\pmb p}+{\pmb Q}_i}^{(i)} \right)
\left( \mN_{{\pmb p}'+ {\pmb Q}_{j}}^{(j)} 
- \theta_{{\pmb p}'+{\pmb Q}_j}^{(j)} \right)
\nonumber \\
+ \sum_{{\pmb p} {\pmb p}' i}
V^{(i)} \left({\pmb p}, {\pmb p'} \right) \,\,
u_{\pmb p}^{(i)} v_{\pmb p}^{(i)} \,
u_{{\pmb p}  '}^{(i)} v_{{\pmb p}  '}^{(i)}\, \,
\nonumber \\
\times \,\, \left(  1- \mF_{{\pmb p}+ {\pmb Q}_{i}}^{(i)}
- \mF_{-{\pmb p}+ {\pmb Q}_{i}}^{(i)}\right) \,
\left(  1-\mF_{{\pmb p}'+ {\pmb Q}_{i}}^{(i)}
-\mF_{-{\pmb p}'+ {\pmb Q}_{i}}^{(i)}\right).
\label{energy1}
\end{gather}
Here $\mF_{{\pmb p}+ {\pmb Q}_{i}}^{(i)}$ 
is the distribution function for Bogoliubov excitations 
with momentum $({\pmb p}+{\pmb Q}_i)$,
\begin{equation}
\mF_{{\pmb p}+ {\pmb Q}_{i}}^{(i)} =
\langle | b_{{\pmb p} +{\pmb Q}_{i} \uparrow}^{(i) \dagger}
b_{{\pmb p} +{\pmb Q}_{i} \uparrow}^{(i)} |\rangle
\,\, = \,\,
\langle | b_{{\pmb p} +{\pmb Q}_{i}\downarrow}^{(i) \dagger}
b_{{\pmb p}+{\pmb Q}_{i} \downarrow}^{(i)} | \rangle,
\label{fp1}
\end{equation}
while $\mN_{{\pmb p}+ {\pmb Q}_{i}}^{(i)}$ is the 
average number of Landau quasiparticles 
in a state $({\pmb p}+{\pmb Q}_i, \sigma)$,
\begin{eqnarray}
\mN_{{\pmb p}+ {\pmb Q}_{i}}^{(i)} &=& \,\,
\langle | a_{{\pmb p} +{\pmb Q}_{i} \uparrow}^{(i) \dagger}
a_{{\pmb p} +{\pmb Q}_{i} \uparrow}^{(i)} |\rangle
\,\, = \,\,
\langle | a_{{\pmb p} +{\pmb Q}_{i}\downarrow}^{(i) \dagger}
a_{{\pmb p}+{\pmb Q}_{i} \downarrow}^{(i)} |\rangle
\nonumber \\
&=& v_{\pmb p}^{(i) \, 2} +
 u_{\pmb p}^{(i) \, 2} \, \mF_{{\pmb p}+ {\pmb Q}_{i}}^{(i)}
- v_{\pmb p}^{(i) \, 2} \, \mF_{-{\pmb p}+ {\pmb Q}_{i}}^{(i)}.
\label{np1}
\end{eqnarray}

If we were in thermodynamic equilibrium 
we could easily find 
the unknown functions
$\mF_{{\pmb p}+ {\pmb Q}_{i}}^{(i)}$ 
and $u_{\pmb p}^{(i)}$ in Eq.\ (\ref{energy1}) 
by requiring minimum of the free energy $F$, 
$F\left[\mF_{{\pmb p}+ {\pmb Q}_{i}}^{(i)}, u_{\pmb p}^{(i)}\right] 
\equiv E- \breve{\mu}_1 n_1 - \breve{\mu}_2 n_2 -T S$,
where the entropy $S[\mF_{{\pmb p}+ {\pmb Q}_{i}}^{(i)}]$ 
is the functional of only $\mF_{{\pmb p}+ {\pmb Q}_{i}}^{(i)}$ 
(see Ref.\ \cite{gh05} for more details).
Since we are {\it not} in thermodynamic equilibrium,
the distribution function for Bogoliubov excitations 
$\mF_{{\pmb p}+ {\pmb Q}_{i}}^{(i)}$ 
in our local analysis should be considered 
as a given `input parameter' 
[it can be found from the corresponding 
Boltzmann kinetic equation (\ref{kineqBog}), 
see Sec.\ IVB].
To determine $u_{\pmb p}^{(i)}$,
we assume that 
even out of equilibrium 
$F$ still has a minimum as the functional of $u_{\pmb p}^{(i)}$
(at fixed $\mF_{{\pmb p}+ {\pmb Q}_{i}}^{(i)}$). 
This assumption, though plausible, 
cannot be proven in our phenomenological approach. 
However, its validity can be justified
by comparison with the results 
of the strict microscopic theory (see Sec.\ IVC).
%

One obtains from the minimization procedure
\begin{equation}
u_{{\pmb p}}^{(i) \, 2} = \frac{1}{2} \,
\left( 1 + \frac{ H_{{\pmb p}+ {\pmb Q}_{i}}^{(i)}+H_{-{\pmb p}+ {\pmb Q}_{i}}^{(i)}}
{2 \mathfrak{E}_{{\pmb p}+ {\pmb Q}_{i}}^{(i)}
+ H_{-{\pmb p}+ {\pmb Q}_{i}}^{(i)}
- H_{{\pmb p}+ {\pmb Q}_{i}}^{(i)}} \right),
\label{up}
\end{equation}
where 
\begin{equation}
H_{\pmb p}^{(i)} = 
\varepsilon_{\pmb p}^{(i)}\left[ \mN_{\pmb p}^{(j)} \right] - \breve{\mu}_i,
\label{localenergy2}
\end{equation}
see Eq.\ (\ref{energy}) for the definition 
of $\varepsilon_{\pmb p}^{(i)}$, and
\begin{equation}
\mathfrak{E}_{{\pmb p}+ {\pmb Q}_{i}}^{(i)}
= \frac{1}{2} \, \left( H_{{\pmb p}+ {\pmb Q}_{i}}^{(i)}
-H_{-{\pmb p}+ {\pmb Q}_{i}}^{(i)}
\right) \, +
\sqrt{\, \frac{1}{4} \, \left( H_{{\pmb p}+ {\pmb Q}_{i}}^{(i)}
+H_{-{\pmb p}+ {\pmb Q}_{i}}^{(i)}\right)^2
+ \mathcal{D}_{\pmb p}^{(i) 2} }
\label{EnergyBog}
\end{equation}
is the energy of a Bogoliubov excitation 
with momentum $({\pmb p}+{\pmb Q}_i)$.
To verify that $\mathfrak{E}_{{\pmb p}+ {\pmb Q}_{i}}^{(i)}$
is indeed the energy,
it is sufficient 
to notice that it is given by the variational derivative of 
the functional $(E-\sum_i \breve{\mu}_i n_i)$ with respect to 
$\mF_{{\pmb p}+ {\pmb Q}_{i}}^{(i)}$,
\begin{equation}
\mathfrak{E}_{{\pmb p}+ {\pmb Q}_{i}}^{(i)} =
\frac{\delta \left(E-\sum_i \breve{\mu}_i n_i \right)}
{\delta \mF_{{\pmb p}+ {\pmb Q}_{i}}^{(i)}}.
\label{variational}
\end{equation}

Finally, $\mathcal{D}_{\pmb p}^{(i)}$ in Eq.\ (\ref{EnergyBog})
is the nonequilibrium energy gap.
It is defined by the equation,
\begin{equation}
\mathcal{D}_{\pmb p}^{(i)} 
= - \sum_{{\pmb p}'} \,V^{(i)} \left({\pmb p}, {\pmb p'} \right) \,
u_{{\pmb p}'}^{(i)} v_{{\pmb p}'}^{(i)}
\, \left(  1-\mF_{{\pmb p}'+ {\pmb Q}_{i}}^{(i)}
-\mF_{-{\pmb p}'+ {\pmb Q}_{i}}^{(i)}\right).
\label{gap1}
\end{equation}
It can be demonstrated that, 
in the linear approximation of Sec.\ II, 
the quantity $\delta \mathcal{D}_{\pmb p}^{(i)} 
\equiv \mathcal{D}_{\pmb p}^{(i)} - \Delta_i$
is related to the integral $O_{{\pmb p}\sigma +}^{(i)}$ 
[see Eq.\ (\ref{O+})], 
\begin{equation}
\delta \mathcal{D}_{\pmb p}^{(i)}= \frac{O_{{\pmb p}\sigma +}^{(i)}}{2}.
\label{dD1}
\end{equation}

Using Eq.\ (\ref{np1}) one can determine 
the nonequilibrium chemical potential $\breve{\mu}_i$
from the requirement that the number density $n_i$ is given by
the sum over all occupied quasiparticle states,
\begin{equation}
n_i = \sum_{{\pmb p}\sigma} 
\mN_{{\pmb p}+ {\pmb Q}_{i}}^{(i)}.
\label{ni2}
\end{equation}

The quantities
$\mN_{{\pmb p}+ {\pmb Q}_{i}}^{(i)}$, 
$\mathfrak{E}_{{\pmb p}+ {\pmb Q}_{i}}^{(i)}$, and
$\mathcal{D}_{\pmb p}^{(i)}$
can be easily found
from, respectively,  Eqs.\ (\ref{np1}), 
(\ref{EnergyBog}), 
and (\ref{gap1}), 
once the distribution 
$\mF_{{\pmb p}+ {\pmb Q}_{i}}^{(i)}$ is specified.
As shown in Ref.\ \cite{gh05}, 
in thermodynamic equilibrium 
the function 
$\mF_{{\pmb p}+ {\pmb Q}_{i}}^{(i)}$
is given by the standard Fermi-Dirac distribution,
\begin{equation}
\widetilde{\mF}_{{\pmb p}+ {\pmb Q}_{i} \, 0}^{(i)} =
\frac{1}{ 1 + {\rm e}^{\mathfrak{E}_{{\pmb p}
+ {\pmb Q}_{i}}^{(i)}/T}}.
\label{FpFp}
\end{equation}
Here and below the equilibrium function
$\mF_{{\pmb p}}^{(i)}$ 
is denoted as 
$\widetilde{\mF}_{{\pmb p} \, 0}^{(i)}$, 
where tilde indicates that
we allow for superfluid currents 
in the system. 
The function $\widetilde{\mF}_{{\pmb p} \, 0}^{(i)}$ 
should not be confused
with the distribution
$\mF_{{\pmb p} \, 0}^{(i)}$ 
[see Sec.\ IIA and, 
in particular, Eq.\ (\ref{fk}) 
for the definition of $\mF_{{\pmb p} \, 0}^{(i)}$].
These functions are equal only 
in the absence of superfluid currents (${\pmb Q}_i=0$),
\begin{equation}
\widetilde{\mF}_{{\pmb p} \, 0}^{(i)} = \mF_{{\pmb p} \, 0}^{(i)}.
\label{F1}
\end{equation}
In this case one also has
for the system in thermodynamic equilibrium
\begin{eqnarray}
\mathfrak{E}_{{\pmb p}}^{(i)} &=& E_{\pmb p}^{(i)},
\label{E1}\\
\mathcal{D}_{\pmb p}^{(i)} &=& \Delta_i.
\label{D1}
\end{eqnarray}
The equalities (\ref{F1})--(\ref{D1}) follow from 
Eqs.\ (\ref{Ek}), (\ref{fk}), and (\ref{gap}) of Sec.\ IIA
and Eqs.\ (\ref{np1})--(\ref{FpFp}) of the present section
(see Ref.\ \cite{gh05} for a detailed derivation).

\subsection{Introducing dynamics} 

The relations between various nonequilibrium quantities discussed above
should be supplemented by the kinetic equation for 
$\mF_{{\pmb p}+ {\pmb Q}_{i}}^{(i)}$, the continuity equation, 
and by the `superfluid' equation, describing the evolution 
of ${\pmb Q}_i$ with time $t$.
The kinetic equation for the Bogoliubov excitations 
takes the standard form,
\begin{equation}
\frac{\partial \mathcal{F}^{(i)}_{ {\pmb p}+{\pmb Q}_i }}{\partial t}
+ \frac{\partial \mathfrak{E}^{(i)}_{{\pmb p}+{\pmb Q}_{i}}}{\partial {\pmb p}} \,
\frac{\partial \mathcal{F}^{(i)}_{ {\pmb p}+{\pmb Q}_i}}{\partial {\pmb r}}
- \frac{\partial \mathfrak{E}^{(i)}_{{\pmb p}+{\pmb Q}_{i}}}{\partial {\pmb r}} \,
\frac{\partial \mathcal{F}^{(i)}_{ {\pmb p}+{\pmb Q}_i }}{\partial {\pmb p}} = 
{\rm St} \left\{ \mathcal{F}^{(j=1, \, 2)}_{{\pmb p}+{\pmb Q}_{j}}\right\}.
\label{kineqBog}
\end{equation}
The collision integral on the right-hand side of this equation
can be easily obtained once the interaction between the excitations
is known (see, e.g., Refs.\ \cite{stephen65,gk69,ag74,wolfle76,aggk81,aggk86,kopnin01}).

The continuity equation is written as
\begin{equation}
\frac{\partial n_{i}}{\partial t} 
+ {\rm div} \,{\pmb j}_{i} =0,
\label{cont}
\end{equation}
where the number density equals
\begin{equation}
n_{i} = \sum_{{\pmb p} \sigma} \mathcal{N}_{{\pmb p}+{\pmb Q}_{i}}
= \sum_{{\pmb p} \sigma} 
\, \left[ v_{\pmb p}^{(i) \, 2} +
 u_{\pmb p}^{(i) \, 2} \, \mathcal{F}_{{\pmb p}+ {\pmb Q}_{i}}^{(i)}
- v_{\pmb p}^{(i) \, 2} 
\, \mathcal{F}_{-{\pmb p}+ {\pmb Q}_{i}}^{(i)} \right],
\label{ni3}
\end{equation}
and the particle current density is given by 
(see, e.g., Ref.\ \cite{leggett65})
\begin{equation}
{\pmb j}_i = \sum_{{\pmb p} \sigma} 
\frac{\partial H_{{\pmb p}+{\pmb Q}_i}^{(i)}}{\partial {\pmb p}}
\, \mathcal{N}_{{\pmb p}+{\pmb Q}_i}^{(i)}.
\label{ji4}
\end{equation}
Since $\mathcal{N}_{{\pmb p}+{\pmb Q}_i}^{(i)}$ differs 
from the equilibrium distribution 
$\mathcal{N}_{{\pmb p} \, 0}^{(i)}$ only in the narrow region near
the Fermi surface, Eq.\ (\ref{ji4}) can be linearized and 
rewritten in the form, similar to Eq.\ (\ref{Ji1}), 
or, after some algebra, 
to Eq.\ (\ref{Ji2}).
Then, using Eqs.\ (\ref{even}) and (\ref{np1}) and noticing that
\begin{equation}
\mathcal{N}_{{\pmb p}+{\pmb Q}_i}^{(i)}-
\mathcal{N}_{-{\pmb p}+{\pmb Q}_i}^{(i)} = 
\mathcal{F}_{{\pmb p}+{\pmb Q}_i}^{(i)} - 
\mathcal{F}_{-{\pmb p}+{\pmb Q}_i}^{(i)},
\label{difference}
\end{equation}
Eq.\ (\ref{ji4}) can be finally presented as
\begin{equation}
{\pmb j}_{i} = \sum_j \, Y_{ij} \, \left[ {\pmb Q}_{j} 
+   \frac{1}{n_j} \, \sum_{{\pmb p} \sigma } \,\,{\pmb p}\, \,
\mathcal{F}_{{\pmb p} + {\pmb Q}_j}^{(j)}  
\right],
\label{ji5}
\end{equation}
where the matrix $Y_{ij}$ is given by Eq.\ (\ref{Yik2}).
For a one-component Fermi-liquid Eq.\ (\ref{ji5}) reduces to
the well-known expression (see, e.g., Refs.\ \cite{ag74,aggk81}),
\begin{equation}
{\pmb j} = \frac{n \, {\pmb Q}}{m} 
+  \sum_{{\pmb p} \sigma } \,\,\frac{\pmb p}{m}\, \,
\mathcal{F}_{{\pmb p} + {\pmb Q}}.  
\label{jone}
\end{equation}
To obtain this formula we employed Eq.\ (\ref{effmass})
and the definitions (\ref{Gij}) and (\ref{Yik2}).

It is important to emphasize that generally 
the mass current density $m_i \,{\pmb j}_i$ 
of $i$-th particle species is not equal to
the momentum density ${\pmb P}_i$, 
\begin{equation}
{\pmb P}_i = \sum_{{\pmb p}\sigma} \left({\pmb p}+{\pmb Q}_i \right) 
\, \mathcal{N}_{{\pmb p}+{\pmb Q}_i}^{(i)} = 
n_i \, {\pmb Q}_i + \sum_{{\pmb p}\sigma} {\pmb p}
\, \mathcal{F}_{{\pmb p}+{\pmb Q}_i}^{(i)}.
\label{momentum}
\end{equation}
However, using Eq.\ (\ref{effmass}) 
one can check that, due to Galilean invariance 
of the system, 
the following equality holds
\begin{equation}
\sum_i m_i \, {\pmb j}_i = \sum_i {\pmb P}_i. 
\label{momentum2}
\end{equation}

Now let us discuss the `superfluid' equation.
It has a natural form 
(see, e.g., Refs.\ \cite{bn69,ag74,aggk81,aggk86}, 
where similar equations 
are written for a one-component liquid),
\begin{equation}
\frac{\partial {\pmb Q}_{i}}{\partial t} 
= - {\pmb \nabla} \breve{\mu}_{i},
\label{sfleq}
\end{equation}
and coincides 
with the `superfluid' equation (4.9) of Ref.\ \cite{aggk81}.
(The authors 
of Ref.\ \cite{aggk81} used different notations.
In particular, our quantity $\breve{\mu}$ 
is related to their invariant potential $\Phi$ by
$\breve{\mu} = \mu- \Phi$, 
where $\mu$ is a constant
from which the authors count energy.)
Eq.\ (\ref{sfleq}) is also equivalent 
to the corresponding equation 
of the Khalatnikov's superfluid hydrodynamics \cite{khalatnikov89},
\begin{equation}
\frac{\partial {\pmb V}_{\rm s}}{\partial t} 
= - {\pmb \nabla} \left({\mu}_{\rm Kh} 
+ \frac{{\pmb V}_{\rm s}^2}{2} \right).
\label{khalat}
\end{equation}
To prove this, we notice that the Khalatnikov's chemical potential 
$\mu_{\rm Kh}$ (per particle mass $m$) is defined
in a reference frame in which ${\pmb V}_{\rm s}={\pmb Q}/m=0$,
while our potential $\breve{\mu}$ 
is defined in the laboratory frame.
They are related by an obvious formula, 
$\breve{\mu} = m \, \left(\mu_{\rm Kh} 
+{\pmb V}_{\rm s}^2/2 \right)$.
Thus, Eqs.\ (\ref{sfleq}) 
and (\ref{khalat}) are indeed equivalent.

Notice that Eq.\ (\ref{sfleq}) will be automatically satisfied 
if we express ${\pmb Q}_i$ and $\breve{\mu}_i$
through the wave function phase $\varphi_i$
of the Cooper-pair condensate 
(see, e.g., Refs.\ \cite{ag74,aggk81,lp80}),
\begin{eqnarray}
{\pmb Q}_i &=& \frac{1}{2} \, {\pmb \nabla} \varphi_i, 
\label{QQ2}\\
-\breve{\mu}_i &=& 
\frac{1}{2} \, \frac{\partial \varphi_i}{\partial t}.
\label{varphi}
\end{eqnarray}
One can verify that, in the linear theory considered in Sec.\ II, 
the phase $\varphi_i$ is related to the `zero harmonic' 
of the function $O_{{\pmb k}\sigma -}^{(i)}$ [see Eq.\ (\ref{O-})]
by the equation
\begin{equation}
\varphi_i = \frac{i}{2 \Delta_i} \, \sum_{{\pmb k}} \,
V_{0}^{(i)} \, \delta s_{{\pmb k}\sigma -}^{(i)}.
\label{phase2}
\end{equation}
All other harmonics of $O_{{\pmb k}\sigma -}^{(i)}$ 
are small and can be neglected
in the quasiclassical limit \cite{bn69}.

\subsection{Inclusion of electromagnetic field 
and comparison with the previous works} 

Eqs.\ (\ref{even}), (\ref{norm}), (\ref{np1})--(\ref{EnergyBog}),
(\ref{gap1}), (\ref{ni2}), (\ref{FpFp}), (\ref{kineqBog}), (\ref{cont}),
(\ref{ji5}), (\ref{momentum}), (\ref{QQ2}), and (\ref{varphi})    
of Secs.\ IVA and IVB fully describe the two-component 
superfluid neutral Fermi-liquid in the limit 
of small ${\pmb q}$ and $\omega$.
The generalization of these equations 
to the case of charged mixtures is straightforward 
(see, e.g., Refs.\ \cite{ag74,aggk81}). 
Namely, they remain essentially the same 
if we redefine the quantities 
${\pmb Q}_i$ and $\breve{\mu}_i$ 
[see Eqs.\ (\ref{QQ2}) and (\ref{varphi})]
through the gauge-invariant combinations,
\begin{eqnarray}
{\pmb Q}_i &=& \frac{1}{2} \, {\pmb \nabla} \varphi_i - e_i \,{\pmb A}, 
\label{QQ3}\\
-\breve{\mu}_i &=& 
\frac{1}{2} \, \frac{\partial \varphi_i}{\partial t} + e_i \, V.
\label{varphi2}
\end{eqnarray}
Notice, however, that the `superfluid' equation 
in the form (\ref{sfleq}) is no longer valid.
As follows from Eqs.\ (\ref{QQ3}) and (\ref{varphi2}),
it should be replaced by
\begin{equation}
\frac{\partial {\pmb Q}_{i}}{\partial t} 
= - {\pmb \nabla} \breve{\mu}_{i} + e_i \, {\pmb E},
\label{sfleq2}
\end{equation}
where ${\pmb E}=-\partial {\pmb A}/\partial t - {\pmb \nabla}\, V$ 
is the self-consistent electric field, which can be found 
with the help of the Maxwell equations.
The right-hand sides of Eqs.\ (\ref{QQ3}) and (\ref{varphi2})
are indeed gauge-invariant since the phase $\varphi_i$ 
transforms as
$\varphi_i \rightarrow \varphi_i - 2 \, e_i \phi$
under the gauge transformation (\ref{gaugeV1}) and (\ref{gaugeA1}) 
(see, e.g., Refs.\ \cite{bn69,lp80}).

The equations formulated above in Sec. IV 
reproduce various limiting cases that were studied in the literature.
First of all, for a {\it one-component} 
Fermi-liquid our equations coincide 
with those obtained by Betbeder-Matibet and Nozieres \cite{bn69} and by 
Aronov and  Gurevich \cite{ag74} 
(see also Refs.\ \cite{aggk81,aggk86}).
Betbeder-Matibet and Nozieres
worked in the linear approximation 
but assumed the most general form 
of the Landau interaction $f({\pmb k}, {\pmb k}')$ 
and pairing potential $V({\pmb k}, {\pmb k}')$.
On the contrary, 
Aronov and Gurevich
derived, from the first principles,
the fully nonlinear system of equations, 
describing the superfluid Fermi-liquid 
in the quasiclassical regime.
However, they completely neglected
the Landau interaction [$f({\pmb k}, {\pmb k}')=0$] 
and took into account only the zero harmonic $V_0$ 
of the pairing interaction ($V_{l}=0$ for $l>0$).

Our equations
for a {\it two-component} Fermi-liquid
were compared only with the results of Sec.\ II 
since we did not find 
a discussion 
in the literature
on transport properties
of strongly interacting superfluid Fermi-mixtures.
We checked that in the linear approximation 
our equations 
reproduce the quasiclassical limit of kinetic equations 
(\ref{kineq1})--(\ref{kineq4}).

\section{Summary} 

This paper discusses transport properties of a mixture of 
two superfluid strongly interacting Fermi-liquids. 
A typical example of such mixture 
is the matter in the internal layers of neutron stars.
To describe the mixture we use the Landau theory of Fermi-liquids
generalized by Larkin and Migdal \cite{lm63,larkin64} 
and by Leggett \cite{leggett65,leggett66} 
to take into account the effects of superfluidity.

Our results are summarized below.

({\it i}) Working in the linear approximation, 
we formulate the system (\ref{kineq1})--(\ref{kineq4}) 
of kinetic equations, 
describing the collisionless superfluid Fermi mixture 
in the self-consistent electromagnetic field.
To derive these equations, 
we follow the approach of Betbeder-Matibet and Nozieres \cite{bn69},
who obtained the kinetic equation for a one-component superfluid Fermi-liquid.
Generally, the system (\ref{kineq1})--(\ref{kineq4}) 
is a straightforward generalization 
of the corresponding equations of Ref.\ \cite{bn69}.
However, there is one nontrivial difference concerning
the form of interaction of Landau quasiparticles 
with the electromagnetic vector potential 
[see Eq.\ (\ref{Lambda_ext})].
For a one-component Fermi-liquid $\alpha_i$ 
in Eq.\ (\ref{Lambda_ext}) is always
equal to electric charge, $\alpha_i=e_i$, 
which is the consequence 
of the Galilean invariance of the system.
On the contrary, for a multi-component mixture $\alpha_i$
is generally given by Eq.\ (\ref{alpha_i}), 
while the Galilean invariance requires only 
that Eq.\ (\ref{effmass}) must be satisfied.

({\it ii}) Using the above kinetic equation,
we determine the particle current density ${\pmb j}_i$ 
of $i$-th particle species
[see Eqs.\ (\ref{Ji}), (\ref{Ji2}), or (\ref{Ji1})]
and show that it is given by the same expression as for 
a nonsuperfluid matter.
For a one-component superfluid Fermi-liquid
this was first shown by Leggett \cite{leggett65}.

({\it iii}) Assuming that the only 
two harmonics $f^{ij}_0$ and $f^{ij}_1$ 
of the Landau quasiparticle interaction 
$f^{ij}({\pmb k}, {\pmb k}')$ are nonzero, 
we calculate the polarization functions 
$P_{00}^{(i)}$, $P_{\rm l}^{(i)}$, and $P_{\rm tr}^{(i)}$ 
[see Eqs.\ (\ref{PlP00}), (\ref{P00}), and (\ref{Ptr})], 
and compare them, in various limiting cases, 
with the results available in the literature
(see, e.g., Refs.\ \cite{leggett65,leggett66,gh05,leinson09}).
We demonstrate, that the functions 
$P_{00, \, {\rm l}, \, {\rm tr}}^{(i)}$
can be expressed through the Landau parameters 
and polarization functions 
$\Pi_{00, \,{\rm l}, \,{\rm tr}}^{(i)}(p_{{\rm F}i}, v_{{\rm F}i})$
of noninteracting Fermi-liquid, 
for which $f^{ij}({\pmb k}, {\pmb k}')=0$.
This result is valid for {\it any} 
smooth pairing potential $V^{(i)}({\pmb k}, {\pmb k}')$ 
and for {\it all} wave vectors ${\pmb q}$ 
and frequencies $\omega$ such that 
$q v_{{\rm F}i}\ll \mu_j$ and $\omega \ll \mu_j$.

({\it iv}) We show that the transverse polarization 
function $P_{\rm tr}^{(i)}$ does not depend on
the Landau quasiparticle interaction 
$f^{ij}({\pmb k}, {\pmb k}')$
in the Pippard limit, 
when $q v_{{\rm F}j} \gg \Delta_i$ 
and $q v_{{\rm F}j} \gg \omega$. 
In this limit it is given by 
$P_{\rm tr}^{(i)}
=\Pi_{{\rm tr}}^{(i)}(p_{{\rm F}i}, v_{{\rm F}i})$.
For 
a one-component Fermi-liquid and $\omega=0$
the same result was obtained previously 
by Leggett \cite{leggett65}.

({\it v}) Finally, 
we formulate a system of nonlinear equations describing 
the nonequilibrium superfluid mixture
in the quasiclassical limit 
($q v_{{\rm F}i} \ll \Delta_j$ 
and $\omega \ll \Delta_j$).
It consists of Eq.\ (\ref{gap1}) for a nonequilibrium energy gap,
scalar kinetic equation (\ref{kineqBog}) for a Bogoliubov excitations,
continuity equation (\ref{cont}), and of Eq.\ (\ref{sfleq2}) 
for the superfluid velocity.
In the linear approximation this system 
is completely equivalent to kinetic equations formulated in Sec.\ II 
and in Ref.\ \cite{bn69}.
Moreover, we verified that 
it reproduces the nonlinear equations 
derived from the first principles 
by Aronov and Gurevich \cite{ag74} 
(see also Refs.\ \cite{aggk81,aggk86}).
To simplify the problem, these authors  
neglected the Landau quasiparticle interaction 
and assumed that the pairing potential is a constant.

The results obtained in this paper 
can be useful
in a variety of applications.
For example, the polarization functions can be used to study 
low-frequency ($\omega \ll \mu_i$) 
long-wavelength ($q v_{{\rm F}i }\ll \mu_j$)
collective modes in superfluid matter of neutron stars 
[see, e.g, recent papers \cite{bd09a,bd09b} 
for an example of such studies in the normal matter].

Also, the complex parts of the polarization functions
$P_{00}^{(i)}$ and $P_{\rm tr}^{(i)}$ determine energy losses
in the Cooper-pairing neutrino emission process.
This process regulates
thermal evolution of neutron stars \cite{yp04,gkyg04,plps04,plps09}
and is especially important in the crust 
of accreting neutron stars,
exhibiting X-ray superbursts \cite{bc09,kzkcbs08}.

Next, the kinetic equation derived in Sec.\ II can be used,
with minor modification, to study the axial-vector response of 
a superfluid Fermi-liquid (see Sec.\ IIIB for more details). 
This problem is also important in application to 
the Cooper-pairing neutrino emission process 
\cite{kr04,kv08,sr09,leinson09}.

One needs
the polarization functions in the Pippard limit
since they describe the plasma screening 
of the interaction between charged particles 
(e.g., protons and electrons) 
in the collision integrals,
determining the kinetic coefficients 
of neutron-star matter \cite{sy07,sy08}.

Finally, the equations presented in Sec.\ IV, 
can be applied to study the response functions 
and collective modes in the equilibrium and in nonequilibrium matter 
(under the condition that $q v_{{\rm F}i}\ll \Delta_j$ and $\omega \ll \Delta_j$).
Furthermore, after specifying the collision integral in Eq.\ (\ref{kineqBog})
(see, e.g., Refs.\ \cite{stephen65,gk69,ag74,wolfle76,aggk81,aggk86,kopnin01}),
the equations of Sec.\ IV 
can be used to calculate the kinetic coefficients 
for a superfluid mixture,
in particular, the thermal conductivity, 
shear and bulk viscosities.
These coefficients are crucial for modeling of
the dynamics of neutron stars \cite{yp04,andersson07}.

While doing this work we had in mind possible applications 
to neutron-star physics. 
However, the results obtained in this paper can 
be applied to any mixture 
of strongly interacting superfluid Fermi-liquids,
for instance, 
to ultracold Fermi-Fermi mixtures,
which have been realized recently 
\cite{tvahd08,wille08etal}.

\section*{Acknowledgments}

The author is very grateful to A.\ D.\ Kaminker and D.\ G.\ Yakovlev
for reading the draft of the paper and valuable comments, 
and to L.\ B.\ Leinson for correspondence.
This research was supported in part
by RFBR (Grant 08-02-00837),
and by the Federal Agency for Science and Innovations
(Grant NSh 2600.2008.2). 
The author also acknowledges
support from the Dynasty Foundation, 
the Mianowski Foundation, 
and from the RF Presidential Program 
(Grant MK--1326.2008.2).

\section*{Appendix A}
\label{App1}

Let us derive Eq.\ (\ref{alpha_i}) 
for the coefficient $\alpha_i$.
This coefficient enters 
the expression (\ref{Lambda_ext})
for the matrix $\Lambda_{{\pmb k}\sigma}^{(i)}$,
describing the interaction of quasiparticles
with the self-consistent electromagnetic field.
Since this matrix is diagonal, it is sufficient 
to consider a mixture 
of strongly interacting {\it normal} Fermi-liquids.
The kinetic equation (\ref{kineq}) then reduces to
\begin{eqnarray}
\omega \, \delta n_{{\pmb k}\sigma}^{(i)}
&=& \left(\xi_{{\pmb k}_+}^{(i)} - \xi_{{\pmb k}_-}^{(i)}\right)
\delta n_{{\pmb k}\sigma}^{(i)} + 
\left(n_{{\pmb k}_- \, 0}^{(i)} - n_{{\pmb k}_+ \, 0}^{(i)}  \right)
\sum_{{\pmb k}' \sigma' j} f^{ij}({\pmb k}, {\pmb k}') 
\, \delta n_{{\pmb k}' \sigma'}^{(j)}
\nonumber\\
&+&\left(n_{{\pmb k}_- \, 0}^{(i)} - n_{{\pmb k}_+ \, 0}^{(i)}  \right)
\left(e_i \, V - \alpha_i \, \frac{{\pmb k}{\pmb A}}{m_i} \right),
\label{kineq2x}
\end{eqnarray}
where $n_{{\pmb k} \, 0}^{(i)}$ is given by Eq.\ (\ref{nk})
and we use the notation 
$\delta n_{{\pmb k}\sigma}^{(i)} \equiv \delta n_{{\pmb k}\sigma \, 11}^{(i)} 
= \left\langle a_{{\pmb k}_- \sigma}^{(i) \dagger} 
a_{{\pmb k}_+ \sigma}^{(i)} \right\rangle$.

The last term in the right-hand side of Eq.\ (\ref{kineq2x}) appears
due to the interaction of Landau quasiparticles with 
the self-consistent electromagnetic field.
The corresponding interaction Hamiltonian has the form
\begin{equation}
H_{\rm em} = \sum_{{\pmb p} \sigma i} 
\left(e_i\, V 
-\alpha_i \, \frac{{\pmb k}{\pmb A}}{m_i} \right) 
a_{{\pmb k}_+ \sigma}^{(i) \dagger} 
a_{{\pmb k}_- \sigma}^{(i)}.
\label{Hem}
\end{equation}
One can easily obtain this term
using $H_{\rm em}$ and an equation 
of motion for the operator 
$a_{{\pmb k}_- \sigma}^{(i) \dagger} 
a_{{\pmb k}_+ \sigma}^{(i)}$ (see, e.g., Ref.\ \cite{bn69}).

In the limit of small ${\pmb q}$, 
which is of our primary interest, 
Eq.\ (\ref{kineq2x}) can be rewritten as
\begin{eqnarray}
\left(\omega - {\pmb q} {\pmb v}_i \right) 
\, \delta n_{{\pmb k}\sigma}^{(i)}
&=& 
-\frac{\partial n_{{\pmb k} \, 0}^{(i)}}{\partial {\pmb k}} {\pmb q} \,
\left[ \sum_{{\pmb k}' \sigma' j} f^{ij}({\pmb k}, {\pmb k}') 
\, \delta n_{{\pmb k}' \sigma'}^{(j)}
+e_i \, V - \alpha_i \, \frac{{\pmb k}{\pmb A}}{m_i} \right].
\label{kineq3x}
\end{eqnarray}
The kinetic equation (\ref{kineq3x}) [or (\ref{kineq2x})]
is not obviously gauge-invariant.
To make the gauge invariance explicit, 
one can notice that ${\pmb k}$ is actually 
a {\it generalized} momentum.
It is related to the {\it real} momentum ${\pmb p}$ 
of a quasiparticle $i$ by
\begin{equation}
{\pmb k} = {\pmb p} + e_i \, {\pmb A}.
\label{kp}
\end{equation}
For a very pedagogical discussion 
of the validity of this expression for Landau quasiparticles, 
see, e.g., Chapter 3, $\S$ 6 of Ref.\ \cite{pn66}.

Thus, we have two momentums, ${\pmb k}$ and ${\pmb p}$,
and our next step will be to express the distribution function 
$\mathcal{N}_{{\pmb p}\sigma}^{(i)} \equiv
n_{{\pmb p}\, 0}^{(i)} + \delta \mathcal{N}_{{\pmb p}\sigma}^{(i)}$ 
for quasiparticles with momentum ${\pmb p}$ through
the distribution function 
$n_{{\pmb k}\, 0}^{(i)}+\delta n_{{\pmb k}\sigma}^{(i)}$
for 
quasiparticles 
with momentum ${\pmb k}$.
Because of the one-to-one correspondence 
between ${\pmb k}$ and ${\pmb p}$,
one has
\begin{equation}
\mathcal{N}_{{\pmb p}\sigma}^{(i)} 
= n_{{\pmb k}\, 0}^{(i)}+\delta n_{{\pmb k}\sigma}^{(i)}.
\label{relation1}
\end{equation}
In view of Eq. (\ref{kp}), in the linear approximation,
\begin{equation}
n_{{\pmb k}\, 0}^{(i)}=n_{{\pmb p}+e_i  {\pmb A} \, 0}^{(i)} 
\approx n_{{\pmb p}\, 0}^{(i)} 
+ \frac{\partial n_{{\pmb p} \, 0}^{(i)}}{\partial {\pmb p}} 
\, e_i \, {\pmb A}.
\label{nknk}
\end{equation}
It follows then from Eq.\ (\ref{relation1})
\begin{equation}
\delta n_{{\pmb k} \sigma}^{(i)} 
=\delta \mathcal{N}_{{\pmb p} \sigma}^{(i)} - 
\frac{\partial n_{{\pmb p}\, 0}^{(i)}}{\partial {\pmb p}} 
\, e_i \, {\pmb A}.
\label{relation2}
\end{equation}
Substituting Eq.\ (\ref{relation2}) 
into the kinetic equation (\ref{kineq3x})
and demanding that the terms which are noninvariant under 
gauge transformations vanish, 
one obtains the condition
\begin{eqnarray}
{\pmb q} {\pmb v}_i
\left[\frac{\partial n_{{\pmb p} \, 0}^{(i)}}{\partial {\pmb p}} e_i \,{\pmb A} \right]
&=& \frac{\partial n_{{\pmb p}\, 0}^{(i)}}{\partial {\pmb p}} {\pmb q}
\left[ \sum_{{\pmb p}' \sigma' j} f^{ij}({\pmb p}, {\pmb p}') 
\, \frac{\partial n_{{\pmb p}' \, 0}^{(j)}}{\partial {\pmb p}'} e_j \, {\pmb A} \,
+ \alpha_i \, \frac{{\pmb p}{\pmb A}}{m_i} \right],
\label{alpha_i2}
\end{eqnarray}
or, after performing an integration, 
the expression (\ref{alpha_i}) for $\alpha_i$.
For a one-component Fermi-liquid Eq.\ (\ref{alpha_i}) 
simplifies with the help of Eq.\ (\ref{effmass}) 
and one gets $\alpha_i=e_i$. 

Using Eq.\ (\ref{alpha_i2}), 
the kinetic equation (\ref{kineq3x})
can be recasted in the well-known gauge-invariant form 
(see, e.g., Ref.\ \cite{pn66}),
\begin{equation}
\left(\omega - {\pmb q} {\pmb v}_i \right) 
\, \delta n_{{\pmb p}\sigma}^{(i)}
=
-\frac{\partial n_{{\pmb p}\, 0}^{(i)}}{\partial {\pmb p}} 
\left[
{\pmb q} \, \sum_{{\pmb p}' \sigma' j} f^{ij}({\pmb p}, {\pmb p}') 
\, \delta n_{{\pmb p}' \sigma'}^{(j)}
+ i \, e_i \,{\pmb E} \right],
\label{kineq4444}
\end{equation}
where ${\pmb E}=i (\omega \, {\pmb A}-{\pmb q} \,V)$ 
is the electric field.

\section*{Appendix B}
\label{App2}

Let us derive Eq.\ (\ref{Aeff1}) directly for a one-component Fermi-liquid.
In this case one has to put $f_0^{ij}=0$ 
and $f_1^{ij}=0$ for $i\neq j$ in all equations.
In what follows we suppress the particle species indices to simplify notations.
We start with Eq.\ (\ref{Aks1}) which can be rewritten as
\begin{equation}
A_{{\pmb k}\sigma} = 
2 e \,\, \frac{{\pmb k} {\pmb A}}{m} 
- \frac{ {\pmb k}}{k_{{\rm F}}^2} \, f_1 \, {\pmb I}
\equiv 2  e \frac{{\pmb k} {\pmb A}_{\rm eff}}{m^\ast},
\label{Aeffeff1}
\end{equation}
where we used the fact that $\alpha=e$ 
for a one-component Fermi-liquid 
(see Appendix A)
and defined
\begin{equation}
{\pmb I} \equiv \sum_{{\pmb k}' \sigma'} {\pmb k}' \, \delta n_{{\pmb k}' \sigma' +}. 
\label{II}
\end{equation}
The quantity ${\pmb I}$ is not gauge-invariant. 
Our aim will be to express ${\pmb I}$ through the vector potential ${\pmb A}$
and the gauge-invariant particle current density ${\pmb j}$, 
which is given by Eq.\ (\ref{Ji2}).
In view of the definition (\ref{II}), 
Eq.\ (\ref{Ji2}) can be rewritten as
\begin{equation}
{\pmb j} = \frac{1}{2} \, 
\left[ \frac{\pmb I}{m^\ast}
- \sum_{{\pmb k} \sigma} 
\frac{\partial \mN_{{\pmb k}\, 0}}{\partial k}  
\,\,  f_1\,\, \frac{ k_\alpha k_\beta}{k_{{\rm F}}^3} \, I_\beta
\right] 
- e \, \frac{n}{m} \, {\pmb A}.
\label{Ji7}
\end{equation}
Here $\alpha$ and $\beta$ are the space indices.
The integral in Eq.\ (\ref{Ji7}) can be easily taken since the function
$\partial \mN_{{\pmb k}\, 0}/{\partial k}$ 
has a sharp maximum near the Fermi-surface. 
As a result, we obtain
\begin{equation}
{\pmb j} = \frac{1}{2} \, 
\left[ \frac{1}{m^\ast}
+\frac{k_{{\rm F}} \, f_1}{3 \pi^2}
\right] \, {\pmb I}
- e \, \frac{n}{m} \, {\pmb A}.
\label{Ji8}
\end{equation}
As follows from Eq.\ (\ref{effmass}) for the effective mass,
the expression in square brackets equals $1/m$.
Thus, one finds
\begin{equation}
{\pmb j} = \frac{1}{2 m}
 \, {\pmb I}
- e \, \frac{n}{m} \, {\pmb A}
\label{Ji9}
\end{equation}
or
\begin{equation}
{\pmb I} = 2 m \left( {\pmb j} +e \frac{n}{m} \, {\pmb A}\right).
\label{III}
\end{equation}
Substituting this formula into Eq.\ (\ref{Aeffeff1}), 
one gets
\begin{equation}
A_{{\pmb k}\sigma} =2 e  
\left( \frac{1}{m} - \frac{n \, f_1}{k_{{\rm F}}^2}\right)
{\pmb k}{\pmb A}
- \frac{2 m f_1}{k_{{\rm F}}^2} \, {\pmb k}{\pmb j}
\equiv 2  e \frac{{\pmb k} {\pmb A}_{\rm eff}}{m^\ast}.
\label{Aeff3}
\end{equation}
Again, using Eq.\ (\ref{effmass}), 
it follows that the expression
in brackets equals $1/m^\ast$.
Eq.\ (\ref{Aeff3}) can then be rewritten as
\begin{equation}
A_{{\pmb k}\sigma} =\frac{2 e}{m^\ast} \,\, {\pmb k}{\pmb A}
- \frac{2 m f_1}{k_{{\rm F}}^2} \, {\pmb k}{\pmb j}
\equiv 2  e \frac{{\pmb k} {\pmb A}_{\rm eff}}{m^\ast}.
\label{Aeff4}
\end{equation}
That is
\begin{equation}
{\pmb A}_{\rm eff}={\pmb A}
- \frac{1}{e}\, \frac{ m m^\ast}{k_{{\rm F}}^2} f_1 \, {\pmb j}.
\label{Aeff5}
\end{equation}
This equation coincides with Eq.\ (\ref{Aeff1})
if the latter is written for a one-component Fermi-liquid.


\end{document}